%% file: NLOSMEFT-handoff.tex
\documentclass[12pt,a4paper]{article}

\input{proleg}

\begin{document}
 

{\it Submitted to: LHC Higgs Cross Section Working Group 2 (Higgs Properties) May 2016.\\  \\}

{\bf The Standard Model Effective Field Theory and 
Next to Leading Order. \\ \\}
{ \centering
Giampiero Passarino$^{1}$, Michael Trott$^{2}$\\
 \centering
[1] Dipartimento di Fisica Teorica, Universita di Torino, INFN, Sezione di Torino, Italy, \\
\centering
[2] Niels Bohr International Academy, University of Copenhagen, Blegdamsvej 17, \\
 \centering
 DK-2100 Copenhagen, Denmark.\\
 }
\renewcommand{\theequation}{\arabic{section}.\arabic{equation}} 

\small{
\section*{Forward} This document was submitted to Working Group 2 of the Higgs Cross Section Working Group.
The derived document in CERN Yellow Report 4 (YR4) \cite{deFlorian:2016spz} differs in a number of key respects from this original version. 
We have highlighted in red in this document some text that is important to consider when judging some content of YR4. 
The highlighted text is not present in the YR4 version of this document. Further comments are made in the Afterword.}
\section{Overview}

In this section we discuss how to interpret data in the Standard Model Effective Field Theory (SMEFT) 
in a transparent manner at leading order (LO) and explain why a next to leading order (NLO)
interpretation of the data is important.\footnote{We thank the following for comments: 
Laure Berthier, N.E.J. Bjerrum-Bohr, 
Cliff Burgess, Mikkel Bj\o rn, Poul Damgaard, Andr\'{e} Mendes , Chris Hays, Gino Isidori, Yun Jiang, 
Fabio Maltoni, Aneesh Manohar, Ben Pecjak, Jose Santiago, Veronica Sanz, William Shepherd, Frank Tackmann. Appearing in acknowledgements
does not imply full or partial endorsement} 
The approach presented  for LO is the one we consider the most simple to enable the
ongoing development of the SMEFT to NLO. The LO approach we present
is written in terms of mass eigenstate fields and is trivially connected to Higgs observables and electroweak precision observables. It can be directly used at LO to interpret the data.

Interpreting the data using theoretical results developed beyond LO (in perturbation theory) can often be crucial to do in the SMEFT. NLO calculations should be used if they are available. 
We discuss the basic issues involved in improving calculations to NLO, and review the advances in this direction that have been achieved to date.
These calculations help characterize (and reduce) theoretical errors of a LO result
and allow the consistent incorporation of precise measurements, such as the LEP pseudo-observables, in the SMEFT.  
NLO interpretations of the data are particularly critical in the event that deviations from the Standard Model (SM) emerge over the course of LHC operations. 
NLO results are being developed in the theoretical community and will become increasingly available over the course of RunII.
Experimental analyses can adopt approaches to LO that will allow these results to be incorporated in the future as efficiently as possible. 

This review provides scientific support for the above statements. The reader who is mostly interested in the LO and NLO summary conclusions can skip directly to the end of this review.

\section{Introduction to the SMEFT}

As exact non-perturbative solutions to 
quantum field theories are rarely known approximate solutions that expand observables perturbatively in a small 
coupling constant or in a ratio of scales are generally developed. Such quantum field theories can be regarded as examples of Effective Field Theory (EFT), 
the treatment of which was pioneered in~\cite{Weinberg:1980wa,Coleman:1969sm,Callan:1969sn}. 
The predictions of the  LO Lagrangian of any EFT are 
approximations of limited applicability and precision.
Developing such predictions beyond leading order is in general extremely useful and straightforward if the LO EFT is well 
defined. 
The ability to improve EFTs from LO to NLO largely explains why they 
have become the standard approach to interpreting data sets of constraints on the SM, as 
reducing theoretical errors to be below experimental errors is required for a precise interpretation of an experimental measurement.

At LHC it is of interest to treat the Standard Model itself as a general EFT.
In this section we briefly outline how the standard straightforward LO formulation of this SMEFT is defined.
We then discuss extending the SMEFT approach to NLO in order to incorporate important QCD and Electroweak 
corrections. 

The SMEFT assumes that $\rm SU(2)_L \times U(1)_Y$ is spontaneously 
broken to $\rm U(1)_{em}$ by the vacuum expectation value of the Higgs field 
($\it v$) and that the observed $J^P =0^+$ scalar is embedded in the Higgs doublet.
The Lagrangian is schematically
\begin{linenomath}
\bea
\mcL_{SMEFT} = \mcL_{SM} + \mcL_{5}+ \mcL_{6} + \mcL_{7} + \mcL_{8} + \cdots
\eea
\end{linenomath}
$\mcL_{5}$ has one operator suppressed by one power of the cut off scale 
($\Lambda$)~\cite{Weinberg:1979sa}. $\mcL_{6}$ has $76$ parameters 
that preserve Baryon number~\cite{Weinberg:1979sa,Buchmuller:1985jz,Abbott:1980zj,Grzadkowski:2010es} in the $N_f = 1$ limit\footnote{Here $N_f$ counts the number of fermion generations.} and four that do not. The baryon preserving operators in $\mcL_{6}$ has  $2499$ parameters in the case
$N_f = 3$ \cite{Alonso:2013hga}.  $\mcL_{7}$ and $\mcL_{8}$ are now known, see Refs. \cite{Lehman:2014jma,Henning:2015alf}.
We label the Wilson coefficients of the operators in $\mcL_5$ as $C_i^5$, operators 
in $\mcL_6$ as $C_i^6$ etc., and have implicitly absorbed the appropriate power of $1/\Lambda$ into the definition of the $C_i$.
When $1/\Lambda$ is made explicit, and pulled out of the Wilson coefficient we will use the tilde superscript as a notation to indicate this, for example $\tilde{C}_i/\Lambda^2$.

\begin{figure}[t]
   \centering
   \includegraphics[width=0.9\textwidth, trim = 30 250 50 80, clip=true]{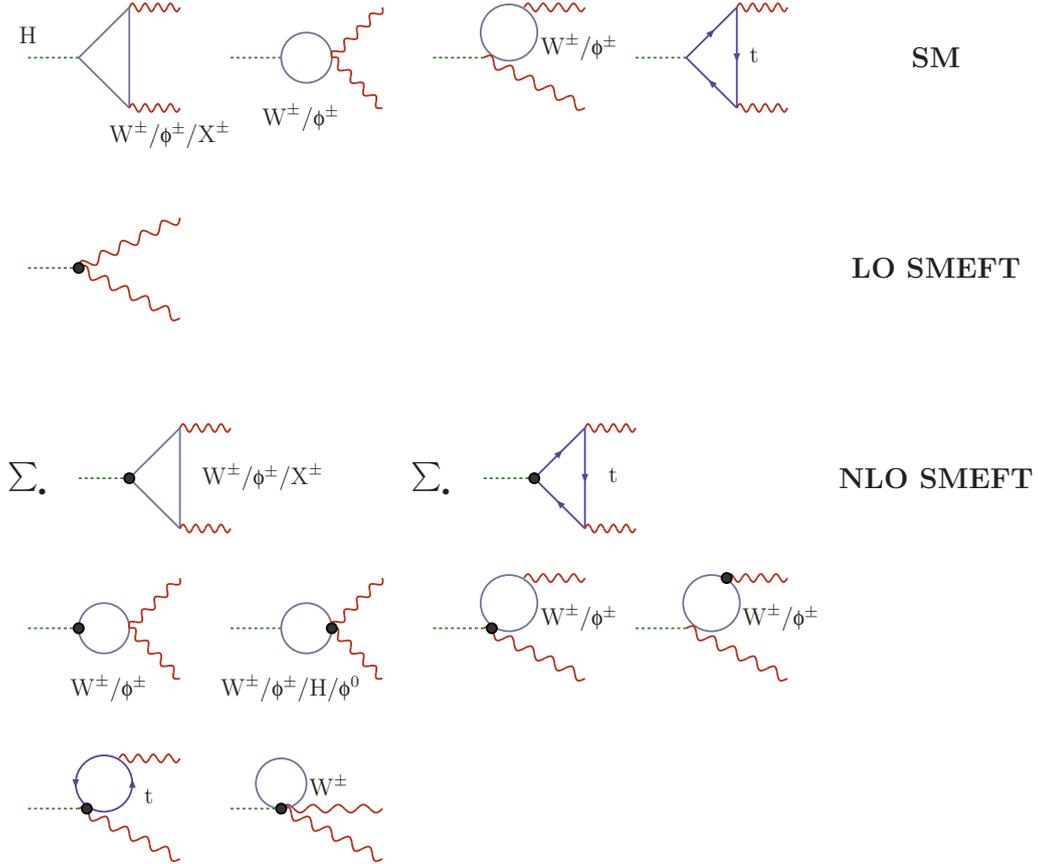}
\caption[]{
Diagrams contributing to the amplitude for
$\PH \to \PGg\PGg$ in the $\mathrm{R}_{\xi}\,$-gauge:
SM (first row), LO SMEFT (second row), and NLO SMEFT.
Black circles denote the insertion of one $\mathcal{L}_6$ operator.
$\sum_{_{\bullet}}$ implies summing over all insertions
in the diagram (vertex by vertex).
For triangles with internal charge flow
($\PQt, \PW^{\pm}, \upphi^{\pm},\PXpm$)
only the clockwise orientation is shown.
Non-equivalent diagrams obtained
by the exchange of the two photon lines are not shown.
Higgs and photon wave-function factors are not included.
The Fadeev-Popov ghost fields are denoted by $\PX$.}
\label{class}
\end{figure}

The SMEFT is a different theory than the SM as it has local contact operators suppressed 
by powers of $1/\Lambda$. To get a feeling for the nature of the LO and NLO predictions in the SMEFT, consider a (lepton number preserving) 
amplitude that can be written as
\begin{linenomath}
\bqa
\mcA &=& \sum_{n=\mrN}^{\infty}\,g_{SM}^n\,\mcA^{(4)}_n +
       \sum_{n=\mrN_6}^{\infty}\,\sum_{l=1}^n\,\sum_{k=1}^{\infty}\,
        g_{SM}^n\,\left[\frac{1}{(\sqrt{2}\,G_{F}\,\Lambda^2)^k}\right]^l\,
        \mcA^{(4+2\,k)}_{n\,l\,k} \spc
\eqa
\end{linenomath}
where $g_{SM}$ is a SM coupling. $G_{F}$ is the Fermi coupling constant and $\Lambda$ is 
again the cut off scale. 
$l$ is an index that indicates the number of SMEFT operator insertions leading to the amplitude, 
and $k$ indicates the inverse mass dimension of the Lagrangian
terms inserted. $N$ is process dependent and  indicates the order of the 
coupling dependence for the leading non-vanishing term in the SM (\eg $N = 1$ for 
$\PH \to \PV\PV$ \etc but $N = 3$ for $\PH \to \PGg\PGg$). $N_6 = N$ for tree 
initiated processes in the SM. For processes that first occur at loop level in the SM,  
$N_6 = N - 2$ as operators in the SMEFT can mediate such decays directly thought a contact 
operator, for example, through a $\mathcal{L}_6$ operator for $\PH \to \PGg\PGg$. For instance, 
the $\PH\PGg\PGg$ (tree) vertex is generated by 
$O_{HB} = H^\dagger \, H \, B^{\mu\nu}\,B_{\mu\nu}$, by 
$O_{HW}^{8}= \  H^\dagger \,B^{\mu\nu}\,B_{\mu \rho} \,\mrD^{\rho}\,\mrD_{\nu}\, H$ \etc
An example of the Feynman diagrams leading to $\mcA$ is given in Fig.~\ref{class}. 

An example of how the SMEFT orders a double expansion in $1/\Lambda$ and the perturbative expansion in SM couplings is as follows.
Consider a tree level $2$ body decay of a single field. The double expansion of such a process 
is given as the following Table~\footnote{Here we have introduced short hand notation where  
$g_{4+2\,k} = 1/(\sqrt{2}\,G_{F}\,\Lambda^2)^k$, so that $g_6$ denotes a single $\mcO^{(6)}$ 
insertion, $g_8$ denotes a single $\mcO^{(8)}$ insertion, $g^2_6$ denotes two, distinct, 
$\mcO^{(6)}$ insertions, etc..}:
\begin{linenomath}
\bea \label{tableNLO}
\begin{array}{llll}
g_{SM}\,/\,\mrdim & \longrightarrow & & \\
\downarrow   & g_{SM}\,\mcA^{(4)}_1  &
               \; + \; g_{SM}\,g_6\,\mcA^{(6)}_{1,1,1} &
               \; + \; g_{SM}\,g_8\,\mcA^{(8)}_{1,1,2} \\
             & g_{SM}^3\,\mcA^{(4)}_3  &
               \; + \; g_{SM}^3\,g_6\,\mcA^{(6)}_{3,1,1} &
               \; + \; g_{SM}^3\,g^2_6\,\mcA^{(6)}_{3,2,1} \\
             &  \dots\dots & \dots\dots & \dots\dots 
\end{array}
\eea
\end{linenomath}
The combination of parameters $g_{SM}\,g_6\,\mcA^{(6)}_{1,1,1}$ defines the LO SMEFT expression 
for the process, including the leading insertion of a higher dimensional operator, and is 
generally well known.
$g_{SM}^3\,g_6\,\mcA^{(6)}_{3,1,1}$ defines the NLO SMEFT amplitude in the perturbative expansion, 
and $g_{SM}\,g_8\,\mcA^{(8)}_{1,1,2}$ defines the NLO SMEFT Lagrangian expansion contribution to the amplitude.
We will refer to these two different NLO effects in this manner in this document.
The discussion here generalizes to cases other than
two body decays of a single field directly. Currently NLO terms in the double expansion present in the SMEFT are generally unknown, in almost every 
process that is of interest phenomenologically. 

The construction of the SMEFT, to all orders, is not based on assumptions on the size of the Wilson coefficients of the higher dimensional operators,
although it does assume that a valid perturbative expansion is present. 
Constructing an NLO SMEFT result means including all operators at a fixed order in the power counting of the theory
or performing a complete one loop calculation for a process, including all of the operators in $\mcL_{6}$ that can contribute. One must add results for real emission (if present) to get a complete
description of a process at NLO in perturbation theory.\footnote{There are different uses of the phrase ``NLO" in the literature. This can refer to a fixed-order NLO
calculation including non-logarithmic terms not fixed by renormalization group evolution, only  an
approximate fixed-order NLO calculation, which includes logarithmic terms
fixed by renormalization group evolution to NLO, and a genuine leading-log calculation, which
uses exact solutions to the RG equations to actually do a resummation. In this work ``NLO in the perturbative expansion" refers to a complete perturbative correction due to SM interactions to the operators in $\mathcal{L}_6$.} 

NLO corrections are a necessary consequence of the SMEFT being a well defined field theory.
The {\it numerical size of the higher order terms} depends upon the high energy (UV) scenario dictating the $\tilde{C}_i$ and $\Lambda$, which is unknown.
Restricting to a particular UV case is not an integral part of a general SMEFT treatment and various cases can be chosen once the general calculation is performed. 
All explicit references to the underlying theory are introduced via the matching procedure in the standard approach to EFTs and power counting, see Refs.\cite{Weinberg:1980wa,Coleman:1969sm,Callan:1969sn,Manohar:1983md,Georgi:1994qn,Kaplan:1995uv,Manohar:1996cq,Cohen:1997rt,Luty:1997fk,Polchinski:1992ed,Rothstein:2003mp,Skiba:2010xn,Burgess:2007pt,Jenkins:2013fya,Jenkins:2013sda,Buchalla:2014eca,Buchalla:2013eza,Gavela:2016bzc} for reviews.
Below we briefly summarize the standard definitions of these terms.

\subsection{Power counting}
The size of corrections to SM results due to $\mcL_{SMEFT}$ interactions are estimated with power counting.\footnote{Differences of opinion about the size of NLO corrections exist in the theory community. Our claim is that any differences of opinion regarding NLO analyses are due to different implicit UV assumptions 
and the data should be reported in a manner that maximizes its potential use in the future, including its use in NLO analyses. This means formalisms that cannot be improved to NLO should be avoided.
We return to this point below.} 
A naive power counting scheme based on the mass dimensions of the operators simply normalizes an operator by the appropriate power of $1/\Lambda$. 
Expansions in $(v/\Lambda)^m$ and $(p^2/\Lambda^2)^m$ are then present, where $p^2$ is a typical invariant momentum flow of a process. Both expansions are relative to the SM interactions.

The Naive Dimensional Analysis (NDA) power counting scheme
incorporates the counting of $\Lambda$ and an estimate of factors of $4 \, \pi$ in the normalization of the operators, see Refs.\cite{Manohar:1983md,Cohen:1997rt,Luty:1997fk,Gavela:2016bzc}
for details. By definition any remaining $4 \, \pi$ dependence, coupling dependence, or alternate scales present in the EFT, can be absorbed into the Wilson coefficients in the matching procedure
if the naive power counting scheme is used.

\subsection{Matching}\label{matchingsection}

Wilson coefficients are determined by calculating  on-shell amplitudes in the UV theory and in the SMEFT and taking
the low energy limit ($E/\Lambda << 1$). The mismatch of the finite terms defines the Wilson coefficient in the matching condition. 

If the value of Wilson coefficients in broad UV scenarios could be inferred in general this would be of significant scientific value.
An example of a scheme that applies to a fairly large set of UV scenarios is the  Artz-Einhorn-Wudka ``potentially-tree-generated''
(PTG) scheme~\cite{Arzt:1994gp,Einhorn:2013kja}. This approach classifies Wilson coefficients for operators in $\mathcal{L}_6$ 
as tree or loop level (suppressed by $g^2/16 \, \pi^2$) essentially using topological matching arguments. This classification scheme corresponds only to a subset of weakly 
coupled and renormalizable UV physics cases, as the topologies considered are (effectively) limited by Lorentz invariance and renormalizability. 
This scheme does not apply to scenarios where any high energy physics is strongly interacting or an EFT itself \cite{Jenkins:2013fya}. 
This scheme should be only considered with caution, as it is not the result of a precise matching calculation.

One can study the Wilson coefficients using dimensional analysis, by restoring $\hbar \neq 1$ in the Lagrangian, as recently discussed in Refs.\cite{Pomarol:2014dya,Panico:2015jxa}.
In this note we do not assume any hierarchy among the couplings discussed in Refs.\cite{Pomarol:2014dya,Panico:2015jxa}, or the claims of these works that
one can unambiguously identify a particular power of hypotheical UV couplings. The reasons we do not adopt these claims is that it is not the case that 
one can unambiguously identify the powers of hypothetical UV couplings present in the $\tilde{C}_i$, as the SM couplings also carry $\hbar$ dimensions and the UV theory is not known.
Further, the matching procedure introduces order one constant terms that can be as large as, or dominant over, any such coupling dependence. In this approach, by performing the calculations without unnecessary
assumptions, it is still possible to study the effect of particular hierarchies and specific UV completions (when they are precisely defined allowing a matching calculation) a posteriori.
It is not necessary or advisable to treat the $\tilde{C}_i$ as anything other than parameters to be constrained by experiment when presenting experimental results.

\subsection{Operator bases for the SMEFT}

The Warsaw basis \cite{Grzadkowski:2010es} for the SMEFT is given in 
Table \ref{warsaw}.
This basis is completely and precisely defined and is fully reduced by the Equations of Motion (EOM). It was the first basis of this form, building upon Ref.\cite{Buchmuller:1985jz}.
No fully reduced basis was present in the literature prior to 2010 when this result was reported. 
The Warsaw basis is the most prominent and standard SMEFT basis in use in the
theoretical community. This is the basis we use to define the straightforward LO approach in subsequent sections.

One can make small field redefinitions of $\mathcal{O}(1/\Lambda^2)$ to shift $\mathcal{L}_{SMEFT}$ by operators proportional to the EOM.
This procedure can be used to eliminate redundant operators from the Lagrangian to obtain a fully reduced basis or to change to a different operator basis.
Operator bases that are related by field redefinitions give equivalent results for physically measured quantities due to the equivalence theorem, see Refs.\cite{Kallosh:1972ap,'tHooft:1972ue,Politzer:1980me}
for the proof of this theorem and its conditions. Field redefinitions are a change of variables in a path integral and do not affect $S$-matrix elements although the source terms in Greens functions can get modified.
If a modification of how $\mathcal{L}_{SMEFT}$ is presented uses manipulations that are not gauge independent field redefinitions, it does not directly satisfy the conditions of the equivalence theorem. 
Any LO Lagrangian construction based on intrinsically gauge dependent manipulations is distinct from a gauge independent operator basis like the Warsaw basis. 
\xxxr{Of particular interest when considering the SMEFT is the question - ``Can one remove all $\mathcal{L}_6$ two derivative interactions of the Higgs field in a gauge independent manner in an operator basis?" 
The answer to this question is no. See section II.B.1 of Ref.\cite{Burgess:2010zq} for a proof on how the corresponding unitary gauge manipulations are mathematically impossible to impose in a gauge independent manner with field redefinitions.} A gauge dependent
construction that removes such interactions would not be referred to as an operator basis in standard EFT literature \cite{Weinberg:1980wa,Coleman:1969sm,Callan:1969sn,Manohar:1983md,Georgi:1994qn,Kaplan:1995uv,Manohar:1996cq,Cohen:1997rt,Luty:1997fk,Polchinski:1992ed,Rothstein:2003mp,Skiba:2010xn,Burgess:2007pt,Jenkins:2013fya,Jenkins:2013sda,Buchalla:2014eca,Buchalla:2013eza,Gavela:2016bzc}.

Any well defined basis can be used.\footnote{Let us consider the general problem of constructing a basis to further clarify this discussion. Every transformation that belongs intrinsically to a specific gauge will inevitably violate the equivalence theorem~\cite{Kallosh:1972ap}.
This theorem can handle non-linear transforms (as long as they are local) but they have to be understood from the point of view of EFT, since (in general) they involve a non trivial Jacobian, i.e. ghost loops (see Ref.~\cite{Bergere:1975tr} and Sect.~2 of Ref.~\cite{Arzt:1993gz}).
Furthermore, any transform that mixes different orders in perturbation theory should be avoided since, inevitably, violation of gauge invariance will be induced. 
Our approach does satisfy all the requests imposed by the equivalence theorem and it is not the aim of this note to provide detailed mathematical evidence for these historically established facts. The interested reader can see Refs.~\cite{Burgess:2010zq,Bergere:1975tr,Arzt:1993gz,Tyutin:2000ht} for more discussion.} There are very few such bases defined in the literature.
We know of the following examples. The Warsaw basis and various constructions that were defined taking  the Warsaw basis and then using the EOM to modify a few terms.\footnote{
Note here we are discussing the relations between such bases that can be obtained with gauge independent field redefinitions to draw a distinction with any ad hoc construction.}
When this is done the resulting construction is generally labelled a ``SILH basis". A version of this later result was reported in Ref.\cite{Contino:2013kra} in 2013.
The different operators that are present in this ``SILH basis" are denoted $\mathcal{O}_i$ and are given by
\begin{linenomath}
\begin{align}
\mathcal{O}_{HW} &=  -i \, g_2 \, (D^\mu H)^\dagger \, \tau^I \, (D^\nu H) \,  W^I_{\mu \, \nu}, &
\mathcal{O}_{HB} &=  -i \, g_1 \, (D^\mu H)^\dagger \,  (D^\nu H) \,  B_{\mu \, \nu}, \\
\mathcal{O}_{W} &=  -\frac{i \, g_2}{2} \, (H^\dagger \,  \overleftrightarrow D^I_\mu H) \,  (D^\nu W^I_{\mu \, \nu}), &
\mathcal{O}_{B} &=  -\frac{i \, g_1}{2} \, (H^\dagger \, \overleftrightarrow{D}^\mu H) \,  (D^\nu B_{\mu \, \nu}),  \\
\mathcal{O}_{T} &=  (H^\dagger \, \overleftrightarrow{D}^\mu H) \,  (H^\dagger \, \overleftrightarrow{D}^\mu H).
\label{SILHops}
\end{align}
\end{linenomath}
We use $Q_i$ for the Warsaw basis operators, $\tau$ is the Pauli matrix, $g_1$ is the $U(1)_Y$ coupling and $g_2$ is the $SU(2)_L$ coupling. See Refs.\cite{Grzadkowski:2010es,Alonso:2013hga}
for more details on notation.
All other operators are the same in these bases.
The  transformation from the Warsaw basis
to the $\mathcal{O}_i$ operators is derived using the SM EOM and found\footnote{
Operator relations of this form were partially discussed in Refs \cite{SanchezColon:1998xg,Kilian:2003xt,Grojean:2006nn} previously.}  to be \cite{Alonso:2013hga}
\begin{linenomath}
\begin{align}
g_1 \, g_2 \, Q_{HWB} &=  4 \, \mathcal{O}_{B} - 4 \, \mathcal{O}_{HB} - 2 \,  \hyp_H \, g_1^2 \, Q_{HB},  \\
g_2^2 \, Q_{HW} &=  4 \, \mathcal{O}_{W} - 4 \, \mathcal{O}_{B} - 4 \, \mathcal{O}_{HW} + 4 \, \mathcal{O}_{HB} + 2 \,  \hyp_H \, g_1^2 \, Q_{HB},  \\
g_1^2 \, \hyp_\ell \, Q_{\substack{H l \\ tt}}^{(1)} &= 2 \, \mathcal{O}_{B} +  \hyp_H \, g_1^2\, \mathcal{O}_T - 
g_1^2\left[\hyp_e Q_{\substack{H e \\ rr}} + \hyp_q Q_{\substack{H q \\ rr}}^{(1)}+\hyp_u Q_{\substack{H u \\ rr}}+ \hyp_d Q_{\substack{H d \\ rr}} \right],  \\
g_2^2  \, Q_{\substack{H l \\ tt}}^{(3)} &=  4 \, \mathcal{O}_{W} - 3 \, g_2^2 \, Q_{H \Box} + 2 \, g_2^2 m_h^2 \, (H^\dagger \, H)^2 - 8 \, g_2^2 \, \lambda \, Q_H - g_2^2 \, Q_{H q}^{(3)},  \\ 
&- 2 \, g_2^2\left( [Y_u^\dagger]_{rr} Q_{\substack{ uH \\ rr}} + [Y_d^\dagger]_{rr} Q_{\substack{dH \\ rr}}+ [Y_e^\dagger]_{rr} Q_{\substack{eH \\ rr}}+h.c. \right).
\label{inversetransform1}
\end{align}
\end{linenomath}
Here the $t$ subscript is a flavour index and the $(1),(3)$ superscripts are operator labels, see Table 1.
In these relations only the flavour singlet component of the operators appears - given by the $tt$ subscript and
the notation $Q_{\substack{H d \\ rr}}$ for the Warsaw basis operators. It is necessary to define what flavour components of the operators are removed and retained in this procedure, as first pointed out in Ref.\cite{Alonso:2013hga}.\footnote{Any attempt to use the ``SILH basis" to describe interactions of vector bosons with fermions, for example in Electroweak Precision Data (EWPD), is not transparent. This basis is subject to the existence of nonintuitive correlations between Wilson coefficients related to EWPD \cite{Trott:2014dma}.} Note that these relationships between operators are not gauge dependent as they follow from gauge independent field redefinitions that satisfy the equivalence theorem.

Higher derivative terms 
are systematically removed using the EOM in favour of other operators without derivatives in the Warsaw basis, as in other EFTs.
This is done for a number of technical reasons and the complete renormalization program for $\mathcal{L}_6$ was only carried out in the Warsaw basis in 
Refs.~\cite{Alonso:2013hga,Grojean:2013kd,Jenkins:2013zja,Jenkins:2013wua} as a result.  
The Warsaw basis and the straightforward LO approach we outline below enables recent NLO work. 
The ``SILH basis" has not been completely renormalized to date. Any LO construction introducing operator normalizations, redefinitions of the SM parameters and EOM manipulations
that are intrinsically gauge dependent is such that the gauge independent results of Ref.~\cite{Alonso:2013hga,Grojean:2013kd,Jenkins:2013zja,Jenkins:2013wua}
cannot be used.
\begin{table}
\begin{center}
\small
\begin{minipage}[t]{4.6cm}
\renewcommand{\arraystretch}{1.5}
\begin{tabular}[t]{c|c}
\multicolumn{2}{c}{$1:X^3$} \\
\hline
$Q_G$                & $f^{ABC} G_\mu^{A\nu} G_\nu^{B\rho} G_\rho^{C\mu} $ \\
$Q_{\widetilde G}$          & $f^{ABC} \widetilde G_\mu^{A\nu} G_\nu^{B\rho} G_\rho^{C\mu} $ \\
$Q_W$                & $\epsilon^{IJK} W_\mu^{I\nu} W_\nu^{J\rho} W_\rho^{K\mu}$ \\ 
$Q_{\widetilde W}$          & $\epsilon^{IJK} \widetilde W_\mu^{I\nu} W_\nu^{J\rho} W_\rho^{K\mu}$ \\
\end{tabular}
\end{minipage}
\begin{minipage}[t]{2.7cm}
\renewcommand{\arraystretch}{1.5}
\begin{tabular}[t]{c|c}
\multicolumn{2}{c}{$2:H^6$} \\
\hline
$Q_H$       & $(H^\dag H)^3$ 
\end{tabular}
\end{minipage}
\begin{minipage}[t]{5.4cm}
\renewcommand{\arraystretch}{1.5}
\begin{tabular}[t]{c|c}
\multicolumn{2}{c}{$3:H^4 D^2$} \\
\hline
$Q_{H\Box}$ & $(H^\dag H)\Box(H^\dag H)$ \\
$Q_{H D}$   & $\ \left(H^\dag D_\mu H\right)^* \left(H^\dag D_\mu H\right)$ 
\end{tabular}
\end{minipage}
\begin{minipage}[t]{2.7cm}

\renewcommand{\arraystretch}{1.5}
\begin{tabular}[t]{c|c}
\multicolumn{2}{c}{$5: \psi^2H^3 + \hbox{h.c.}$} \\
\hline
$Q_{eH}$           & $(H^\dag H)(\bar l_p e_r H)$ \\
$Q_{uH}$          & $(H^\dag H)(\bar q_p u_r \widetilde H )$ \\
$Q_{dH}$           & $(H^\dag H)(\bar q_p d_r H)$\\
\end{tabular}
\end{minipage}

\vspace{0.25cm}

\begin{minipage}[t]{4.7cm}
\renewcommand{\arraystretch}{1.5}
\begin{tabular}[t]{c|c}
\multicolumn{2}{c}{$4:X^2H^2$} \\
\hline
$Q_{H G}$     & $H^\dag H\, G^A_{\mu\nu} G^{A\mu\nu}$ \\
$Q_{H\widetilde G}$         & $H^\dag H\, \widetilde G^A_{\mu\nu} G^{A\mu\nu}$ \\
$Q_{H W}$     & $H^\dag H\, W^I_{\mu\nu} W^{I\mu\nu}$ \\
$Q_{H\widetilde W}$         & $H^\dag H\, \widetilde W^I_{\mu\nu} W^{I\mu\nu}$ \\
$Q_{H B}$     & $ H^\dag H\, B_{\mu\nu} B^{\mu\nu}$ \\
$Q_{H\widetilde B}$         & $H^\dag H\, \widetilde B_{\mu\nu} B^{\mu\nu}$ \\
$Q_{H WB}$     & $ H^\dag \tau^I H\, W^I_{\mu\nu} B^{\mu\nu}$ \\
$Q_{H\widetilde W B}$         & $H^\dag \tau^I H\, \widetilde W^I_{\mu\nu} B^{\mu\nu}$ 
\end{tabular}
\end{minipage}
\begin{minipage}[t]{5.2cm}
\renewcommand{\arraystretch}{1.5}
\begin{tabular}[t]{c|c}
\multicolumn{2}{c}{$6:\psi^2 XH+\hbox{h.c.}$} \\
\hline
$Q_{eW}$      & $(\bar l_p \sigma^{\mu\nu} e_r) \tau^I H W_{\mu\nu}^I$ \\
$Q_{eB}$        & $(\bar l_p \sigma^{\mu\nu} e_r) H B_{\mu\nu}$ \\
$Q_{uG}$        & $(\bar q_p \sigma^{\mu\nu} T^A u_r) \widetilde H \, G_{\mu\nu}^A$ \\
$Q_{uW}$        & $(\bar q_p \sigma^{\mu\nu} u_r) \tau^I \widetilde H \, W_{\mu\nu}^I$ \\
$Q_{uB}$        & $(\bar q_p \sigma^{\mu\nu} u_r) \widetilde H \, B_{\mu\nu}$ \\
$Q_{dG}$        & $(\bar q_p \sigma^{\mu\nu} T^A d_r) H\, G_{\mu\nu}^A$ \\
$Q_{dW}$         & $(\bar q_p \sigma^{\mu\nu} d_r) \tau^I H\, W_{\mu\nu}^I$ \\
$Q_{dB}$        & $(\bar q_p \sigma^{\mu\nu} d_r) H\, B_{\mu\nu}$ 
\end{tabular}
\end{minipage}
\begin{minipage}[t]{5.4cm}
\renewcommand{\arraystretch}{1.5}
\begin{tabular}[t]{c|c}
\multicolumn{2}{c}{$7:\psi^2H^2 D$} \\
\hline
$Q_{H l}^{(1)}$      & $(H^\dag i\overleftrightarrow{D}_\mu H)(\bar l_p \gamma^\mu l_r)$\\
$Q_{H l}^{(3)}$      & $(H^\dag i\overleftrightarrow{D}^I_\mu H)(\bar l_p \tau^I \gamma^\mu l_r)$\\
$Q_{H e}$            & $(H^\dag i\overleftrightarrow{D}_\mu H)(\bar e_p \gamma^\mu e_r)$\\
$Q_{H q}^{(1)}$      & $(H^\dag i\overleftrightarrow{D}_\mu H)(\bar q_p \gamma^\mu q_r)$\\
$Q_{H q}^{(3)}$      & $(H^\dag i\overleftrightarrow{D}^I_\mu H)(\bar q_p \tau^I \gamma^\mu q_r)$\\
$Q_{H u}$            & $(H^\dag i\overleftrightarrow{D}_\mu H)(\bar u_p \gamma^\mu u_r)$\\
$Q_{H d}$            & $(H^\dag i\overleftrightarrow{D}_\mu H)(\bar d_p \gamma^\mu d_r)$\\
$Q_{H u d}$ + h.c.   & $i(\widetilde H ^\dag D_\mu H)(\bar u_p \gamma^\mu d_r)$\\
\end{tabular}
\end{minipage}

\vspace{0.25cm}

\begin{minipage}[t]{4.95cm}
\renewcommand{\arraystretch}{1.5}
\begin{tabular}[t]{c|c}
\multicolumn{2}{c}{$8:(\bar LL)(\bar LL)$} \\
\hline
$Q_{ll}$        & $(\bar l_p \gamma_\mu l_r)(\bar l_s \gamma^\mu l_t)$ \\
$Q_{qq}^{(1)}$  & $(\bar q_p \gamma_\mu q_r)(\bar q_s \gamma^\mu q_t)$ \\
$Q_{qq}^{(3)}$  & $(\bar q_p \gamma_\mu \tau^I q_r)(\bar q_s \gamma^\mu \tau^I q_t)$ \\
$Q_{lq}^{(1)}$                & $(\bar l_p \gamma_\mu l_r)(\bar q_s \gamma^\mu q_t)$ \\
$Q_{lq}^{(3)}$                & $(\bar l_p \gamma_\mu \tau^I l_r)(\bar q_s \gamma^\mu \tau^I q_t)$ 
\end{tabular}
\end{minipage}
\begin{minipage}[t]{5.45cm}
\renewcommand{\arraystretch}{1.5}
\begin{tabular}[t]{c|c}
\multicolumn{2}{c}{$8:(\bar RR)(\bar RR)$} \\
\hline
$Q_{ee}$               & $(\bar e_p \gamma_\mu e_r)(\bar e_s \gamma^\mu e_t)$ \\
$Q_{uu}$        & $(\bar u_p \gamma_\mu u_r)(\bar u_s \gamma^\mu u_t)$ \\
$Q_{dd}$        & $(\bar d_p \gamma_\mu d_r)(\bar d_s \gamma^\mu d_t)$ \\
$Q_{eu}$                      & $(\bar e_p \gamma_\mu e_r)(\bar u_s \gamma^\mu u_t)$ \\
$Q_{ed}$                      & $(\bar e_p \gamma_\mu e_r)(\bar d_s\gamma^\mu d_t)$ \\
$Q_{ud}^{(1)}$                & $(\bar u_p \gamma_\mu u_r)(\bar d_s \gamma^\mu d_t)$ \\
$Q_{ud}^{(8)}$                & $(\bar u_p \gamma_\mu T^A u_r)(\bar d_s \gamma^\mu T^A d_t)$ \\
\end{tabular}
\end{minipage}
\begin{minipage}[t]{4.75cm}
\renewcommand{\arraystretch}{1.5}
\begin{tabular}[t]{c|c}
\multicolumn{2}{c}{$8:(\bar LL)(\bar RR)$} \\
\hline
$Q_{le}$               & $(\bar l_p \gamma_\mu l_r)(\bar e_s \gamma^\mu e_t)$ \\
$Q_{lu}$               & $(\bar l_p \gamma_\mu l_r)(\bar u_s \gamma^\mu u_t)$ \\
$Q_{ld}$               & $(\bar l_p \gamma_\mu l_r)(\bar d_s \gamma^\mu d_t)$ \\
$Q_{qe}$               & $(\bar q_p \gamma_\mu q_r)(\bar e_s \gamma^\mu e_t)$ \\
$Q_{qu}^{(1)}$         & $(\bar q_p \gamma_\mu q_r)(\bar u_s \gamma^\mu u_t)$ \\ 
$Q_{qu}^{(8)}$         & $(\bar q_p \gamma_\mu T^A q_r)(\bar u_s \gamma^\mu T^A u_t)$ \\ 
$Q_{qd}^{(1)}$ & $(\bar q_p \gamma_\mu q_r)(\bar d_s \gamma^\mu d_t)$ \\
$Q_{qd}^{(8)}$ & $(\bar q_p \gamma_\mu T^A q_r)(\bar d_s \gamma^\mu T^A d_t)$\\
\end{tabular}
\end{minipage}

\vspace{0.25cm}

\begin{minipage}[t]{3.75cm}
\renewcommand{\arraystretch}{1.5}
\begin{tabular}[t]{c|c}
\multicolumn{2}{c}{$8:(\bar LR)(\bar RL)+\hbox{h.c.}$} \\
\hline
$Q_{ledq}$ & $(\bar l_p^j e_r)(\bar d_s q_{tj})$ 
\end{tabular}
\end{minipage}
\begin{minipage}[t]{5.5cm}
\renewcommand{\arraystretch}{1.5}
\begin{tabular}[t]{c|c}
\multicolumn{2}{c}{$8:(\bar LR)(\bar L R)+\hbox{h.c.}$} \\
\hline
$Q_{quqd}^{(1)}$ & $(\bar q_p^j u_r) \epsilon_{jk} (\bar q_s^k d_t)$ \\
$Q_{quqd}^{(8)}$ & $(\bar q_p^j T^A u_r) \epsilon_{jk} (\bar q_s^k T^A d_t)$ \\
$Q_{lequ}^{(1)}$ & $(\bar l_p^j e_r) \epsilon_{jk} (\bar q_s^k u_t)$ \\
$Q_{lequ}^{(3)}$ & $(\bar l_p^j \sigma_{\mu\nu} e_r) \epsilon_{jk} (\bar q_s^k \sigma^{\mu\nu} u_t)$
\end{tabular}
\end{minipage}
\end{center}
\caption{\label{warsaw}
The $\mathcal{L}_6$ operators built from Standard Model fields which conserve baryon number in the Warsaw basis 
\cite{Grzadkowski:2010es}. The flavour labels of the form $p,r,s,t$ on the $Q$ operators are suppressed on the left hand side of
the tables.}
\end{table}

\subsection{Rotating to mass eigenstate fields}\label{rotation}
Expanding around the vev in unitary gauge and rotating to mass eigenstate fields, the LO modification of the SM interactions in the SMEFT come about
in a straightforward manner.\footnote{The operator basis for the SMEFT remains the Warsaw basis when the interaction terms are expanded in terms of
mass eigenstate fields in unitary gauge. Operator bases are gauge independent, satisfy the equivalence theorem, and do not change when the SMEFT is improved from LO to NLO.}
Here we list the most phenomenologically relevant terms present for mass eigenstate fields, the remaining interactions unlisted come from Class $1,3,5,6,8$ operators in Table 1.
It is not required in our approach to specify all interactions as we make no assertion that these mass eigenstate interactions listed are an operator basis. 
As the theory should be canonically normalized,
we denote coupling parameters in the canonically normalized SMEFT with bar superscripts. This use of bar notation is distinct from
bar superscripts on fermion fields where $\bar{\psi} = \psi^\dagger \gamma^0$.
The following section is largely taken from Ref. \cite{Alonso:2013hga}.

\subsubsection{SM Lagrangian}
We define the SM Lagrangian as
\begin{linenomath}
\begin{align}
\mathcal{L}_{\rm SM} &= -\frac14 G_{\mu \nu}^A G^{A\mu \nu}-\frac14 W_{\mu \nu}^I W^{I \mu \nu} -\frac14 B_{\mu \nu} B^{\mu \nu}
+ (D_\mu H^\dagger)(D^\mu H)
+\sum_{\psi=q,u,d,l,e} \overline \psi\, i \slashed{D} \, \psi\nn
&-\lambda \left(H^\dagger H -\frac12 v^2\right)^2- \biggl[ H^{\dagger j} \overline d\, Y_d\, q_{j} + \widetilde H^{\dagger j} \overline u\, Y_u\, q_{j} + H^{\dagger j} \overline e\, Y_e\,  l_{j} + \hbox{h.c.}\biggr],
\label{sm}
\end{align}
\end{linenomath}
which implicitly defines most of our notational conventions. Note $\tilde{H}^j = \epsilon_{jk} \, H^{\dagger k}$. We have suppressed reference to the 
$\tilde{\theta}$ gauge dual operators of the form $\tilde{\theta} \, F^{\mu \, \nu} \, \tilde{F}_{\mu \, \nu}$. These terms are known to be experimentally small. For dual gauge fields we use the convention $\tilde{F}_{\mu \, \nu} = \epsilon_{\mu \, \nu \, \alpha \, \beta} \, F^{\alpha \, \beta}$ 
with $\epsilon_{0123} = + 1$. See Ref \cite{Alonso:2013hga,Jenkins:2013zja,Jenkins:2013wua} for more details on notation.

\subsubsection{Higgs mass and self-couplings}\label{sec:Hmass}
The potential in the SMEFT is
\begin{linenomath}
\begin{align}
V(H) &= \lambda \left(H^\dagger H -\frac12 v^2\right)^2 - C_H \left( H^\dagger H \right)^3,
\label{pot}
\end{align}
\end{linenomath}
yielding the new minimum
\begin{linenomath}
\begin{align}
\langle H^\dagger H \rangle &= \frac{v^2}{2} \left( 1+ \frac{3 C_H v^2}{4 \lambda} \right) \equiv \frac12 v_T^2.
\end{align}
\end{linenomath}
The scalar field can be written in unitary gauge as
\begin{linenomath}
\begin{align}
H &= \frac{1}{\sqrt 2} \left(\begin{array}{c}
0 \\
 \left[ 1+ \ckin \right]  h + v_T
 \end{array}\right),
 \label{Hvev}
\end{align}
\end{linenomath}
where
\begin{linenomath}
\begin{align}\label{chkindef}
\ckin &\equiv \left(C_{H\Box}-\frac14 C_{HD}\right)v^2, &
v_T &\equiv \left( 1+ \frac{3 C_H v^2}{8 \lambda} \right) v.
\end{align}
\end{linenomath}
The coefficient of $h$ in Eq.~(\ref{Hvev}) is no longer unity, in order for the Higgs boson kinetic term to be properly normalized when the dimension-six operators are included. In what follows
we can exchange $v_T$ for $v$ when this parameter multiplies a operator in $\mathcal{L}_6$ as the difference is NLO in the lagrangian expansion.The kinetic terms 
\begin{linenomath}
\begin{align}
{\cal L}^{(6)} &= (D_\mu H^\dagger)(D^\mu H) + C_{H \Box} \left( H^\dagger H \right) \Box \left( H^\dagger H \right) + C_{HD} \left( H^\dagger D^\mu H \right)^* \left( H^\dagger D_\mu H \right),
\end{align}
\end{linenomath}
and the potential in Eq.~(\ref{pot}) yield
\begin{linenomath}
\begin{align}
{\cal L}^{(6)} &= {1 \over 2} \left( \partial_\mu h \right)^2 -\ckin \left[ \frac{h^2}{v^2} +2 \,  \frac{h}{v} \right](\partial_\mu h)^2 - \frac{m_h^2}{2} h^2 -\lambda v_T \left(1-\frac{5 C_H v^2}{2 \lambda} +3 \ckin  \right)h^3 \nn
&-\frac14 \lambda  \left(1-\frac{15 C_H v^2}{2 \lambda} +4 \ckin \right)h^4 +\frac34 C_H v h^5+\frac18 C_H h^6,
\end{align}
\end{linenomath}
for the $h$ self-interactions. The Higgs boson mass is
\begin{linenomath}
\begin{align}
m_h^2 &= 2 \lambda v_T^2 \left(1-\frac{3 C_H v^2}{2 \lambda} + 2 \ckin \right)\,.
\end{align}
\end{linenomath}

\subsubsection{Yukawa couplings}\label{sec:yuk}

The Lagrangian terms in the unbroken theory
\begin{linenomath}
\begin{align}
{\cal L} &= - \biggl[ H^{\dagger j} \overline d_r\, \left[Y_d \right]_{rs}\, q_{j s} + \widetilde H^{\dagger j} \overline u_r\, \left[ Y_u \right]_{rs} \, q_{j s} 
+ H^{\dagger j} \overline e_r\, \left[Y_e\right]_{rs}\,  l_{j s} + \hbox{h.c.}\biggr] \\
&+ \left[ C^*_{\substack{dH \\ sr}} \left( H^\dagger H \right) H^{\dagger j}  \overline d_r q_{j s} + C^*_{\substack{uH \\ sr}} \left( H^\dagger H \right) \tilde H^{\dagger j}  \overline u_r q_{j s} + C^*_{\substack{eH \\ sr}} \left( H^\dagger H \right) H^{\dagger j}  \overline e_r l_{j s} + \hbox{h.c.} \right], \nonumber
\end{align}
\end{linenomath}
yield the fermion mass matrices
\begin{linenomath}
\begin{align}
\left[ M_\psi \right]_{rs} &= \frac{v_T}{\sqrt 2} \left( \left[Y_\psi \right]_{rs}   - \frac12 v^2 C^*_{\substack{\psi H \\ sr}}   \right), \qquad \psi=u,d,e
\end{align}
\end{linenomath}
in the broken theory.  The  coupling matrices of the $h$ boson to the fermions $\mathcal{L}=- h \, \overline u \, \mathcal{Y}\, q + h.c + \ldots$ are
\begin{linenomath}
\begin{align}
\left[ {\cal Y}_\psi \right]_{rs} &= \frac{1}{\sqrt 2}  \left[ Y_\psi \right]_{rs}\left[ 1+ \ckin \right]  - \frac{3}{2 \, \sqrt{2}} v^2 C^*_{\substack{\psi H \\ sr}} 
\nn
& = \frac{1}{v_T}\left[ M_\psi \right]_{rs} \left[ 1+ \ckin  \right]   - \frac{v^2}{\sqrt{2}} C^*_{\substack{\psi H \\ sr}} , 
\qquad \psi=u,d,e.
\label{5.12}
\end{align}
\end{linenomath}
The fermion fields can be rotating to diagonal mass eigenstates with $3 \times 3$ unitary matricies
$\mathcal{U}$ as
\begin{linenomath}
\begin{align}
\psi_L &= \mathcal{U}(\psi,L) \, \psi_L', & \psi_R &= \mathcal{U}(\psi,R) \, \psi_R',
\end{align}
\end{linenomath}
where the measured masses $\hat{m}^i_\psi$ are
\begin{linenomath}
\bea
\mathcal{U}^\dagger(\psi,R)  \, \left[ M_\psi \right] \, \mathcal{U}(\psi,L) = \delta_{ij} \, \hat{m}^i_\psi, \quad \quad 
\begin{array}{c} i = \{u,c,t \}, \quad \psi =u, \\ i = \{d,s,b \}, \quad \psi =d, \\ i = \{e,u,\tau \}, \quad \psi =e. \end{array}
\eea
\end{linenomath}
For the complex Yukawa coupling the higgs to the mass eigenstate fermion fields 
\begin{linenomath}
\bea
\mathcal{L}=- h \, \left[ {\cal Y}_\psi \right]_{rs} \, \bar{\psi}_r \, P_L \,  \psi_s + h.c. 
\eea
\end{linenomath}
where $P_L =  (1- \gamma_5)/2$ and
\begin{linenomath}
\bea
\left[ {\cal Y}_\psi \right]_{rs} =  \delta_{rs} \, \frac{\hat{m}^r_\psi}{v_T} \,  \left[ 1+ \ckin  \right]  - \frac{v^2}{\sqrt{2}} \left[\mathcal{U}^\dagger(\psi,R) \, C^*_{\substack{\psi H}}  \, \mathcal{U}(\psi,L) \right]_{rs}.
\eea
\end{linenomath}
The Yukawa matricies are off diagonal in general and not simply proportional to the fermion mass matrices as in the SM, as indicated by the second term.
The CKM and PMNS matrices control flavour violating interactions in the SM and are defined as 
\begin{linenomath}
\bea
V_{CKM} =  \mathcal{U}(u,L)^\dagger \, \mathcal{U}(d,L), \quad \quad U_{PMNS} = \mathcal{U}(e,L)^\dagger \, \mathcal{U}(\nu,L),
\eea
\end{linenomath}
when the $\mathcal{U}$ matricies only rotate between the weak and mass eigenstates in the SM.
The definition of these matricies in the SMEFT is a convention choice. Here we choose to define these matricies so that the masses are taken to diagonal form including the $\mathcal{L}_6$ interactions.

\subsubsection{Gauge boson masses and couplings}\label{sec:Gmass} 
The relevant CP even $\mathcal{L}_6$ terms are
\begin{linenomath}
\begin{align}
\lsix &=  C_{HG} H^\dagger H G_{\mu \nu}^A G^{A\mu \nu} + C_{HW} H^\dagger H W_{\mu \nu}^I W^{I \mu \nu}  + C_{HB} H^\dagger H B_{\mu \nu} B^{\mu \nu}  \nn
&+ C_{HWB} H^\dagger \tau^I H W^I_{\mu \nu} B^{\mu \nu} + C_G f^{ABC} G^{A \nu}_\mu G^{B \rho}_\nu G^{C \mu}_\rho + C_W \epsilon^{IJK} W^{I \nu}_\mu W^{J \rho}_\nu W^{K \mu}_\rho \ .
\end{align}
\end{linenomath}
The gauge fields need to be redefined, so that the kinetic terms are properly normalized and diagonal. The first step is to redefine the gauge fields
\begin{linenomath}
\begin{align}
G_\mu^A &= \mathcal{G}_\mu^A \left(1 + C_{HG} v_T^2 \right), &
W^I_\mu  &=  \mathcal{W}^I_\mu \left(1 + C_{HW} v_T^2 \right), &
B_\mu  &=  \mathcal{B}_\mu \left(1 + C_{HB} v_T^2 \right).
\label{5.16a}
\end{align}
\end{linenomath}
The modified coupling constants are
\begin{linenomath}
\begin{align}
\gcg &= g_3 \left(1 + C_{HG} \, v_T^2 \right), & \gcw &= g_2 \left(1 + C_{HW} \, v_T^2 \right), & \gcb &= g_1 \left(1 + C_{HB} \, v_T^2 \right),
\label{5.16b}
\end{align}
\end{linenomath}
so that the products $g_3 G_\mu^A=\gcg \mathcal{G}_\mu^A$, etc.\ are unchanged. The mass eigenstate basis is given by~\cite{Grinstein:1991cd}
\begin{linenomath}
\begin{align}
\left[ \begin{array}{cc}  \mathcal{W}_\mu^3 \\ \mathcal{B}_\mu \end{array} \right]
&=  
\left[ \begin{array}{cc}  1   &  -  \frac{1}{2} \,  v_T^2 \,  C_{HWB} \\
- \frac{1}{2} \,  v_T^2 \,  C_{HWB} & 1 \end{array} \right] \, \left[ \begin{array}{cc} \cos \tc  &  \sin \tc \\
-\sin \tc & \cos \tc \end{array} \right] \left[ \begin{array}{cc}  \mathcal{Z}_\mu \\ \mathcal{A}_\mu \end{array} \right], 
\end{align}
\end{linenomath}
where the rotation angle is
\begin{linenomath}
\begin{align}
\tan \tc &= \frac{\gcb}{\gcw} + \frac{v_T^2}{2}  \,  C_{HWB} \, \left[1 -  \frac{\gcb^2}{\gcw^2}\right].
\end{align}
\end{linenomath}
The $W$ and $Z$ masses are
\begin{linenomath}
\begin{align}
\bar{M}_W^2 &= \frac{\gcw^2 v_T^2}{4} , \nn
\bar{M}_Z^2 &= \frac{v_T^2}{4}(\gcb^2+\gcw^2)+\frac{1}{8}v_T^4 C_{HD} (\gcb^2+\gcw^2)+\frac{1}{2}v_T^4 \gcb\gcw C_{HWB}.
\end{align}
\end{linenomath}
The covariant derivative is 
\begin{linenomath}
\begin{align}
D_{\mu} = \partial_\mu + i \, \frac{\gcw}{\sqrt{2}} \left[\mathcal{W}_{\mu}^+ T^+ + \mathcal{W}_{\mu}^- T^- \right] + i \bar{g}_Z \left[T_3 - \sc^2 Q \right] \mathcal{Z}_\mu + 
i \, \ec \, Q \, \mathcal{A}_\mu,
\end{align}
\end{linenomath}
where $Q=T_3+Y$,  and the effective couplings are given by
\begin{linenomath}
\begin{align}
\ec &= \frac{\gcb \gcw}{\sqrt{\gcw^2+\gcb^2}} \left[ 1 - \frac{\gcb \gcw}{\gcw^2+\gcb^2} v_T^2 C_{HWB} \right] = \gcw \, \sin \tc - \frac{1}{2} \, \cos \tc  \, \gcw \, v_T^2 \, C_{HWB}, \nn
\gcZ &= \sqrt{\gcw^2 + \gcb^2} + \frac{\gcb \gcw}{\sqrt{\gcw^2 + \gcb^2} } v_T^2  C_{HWB}=  \frac{\ec}{\sin \tc \cos \tc}  \left[1 +  \frac{\gcb^2+\gcw^2}{2 \gcb \gcw} v_T^2C_{HWB}\right] ,\nn
\sc^2 &= \sin^2 \tc =  \frac{\gcb^2}{{\gcw^2 + \gcb^2}} + \frac{\gcb \gcw (\gcw^2-\gcb^2)}{(\gcb^2+\gcw^2)^2}  v_T^2 C_{HWB} .
\end{align}
\end{linenomath}
The relevant CP odd $\mathcal{L}^{\cancel{CP}}_6$ terms are
\begin{linenomath}
\begin{align}
\mathcal{L}^{\cancel{CP}}_6 &=  C_{H\tilde{G}} H^\dagger H \tilde{G}_{\mu \nu}^A G^{A\mu \nu} + C_{H\tilde{W}} H^\dagger H \tilde{W}_{\mu \nu}^I W^{I \mu \nu}  + C_{H \tilde{B}} H^\dagger H \tilde{B}_{\mu \nu} B^{\mu \nu}  \nn
&+ C_{H \tilde{W} B} H^\dagger \tau^I H \tilde{W}^I_{\mu \nu} B^{\mu \nu} + C_{\tilde{G}} f^{ABC} \tilde{G}^{A \nu}_\mu G^{B \rho}_\nu G^{C \mu}_\rho + C_{\tilde{W}} \epsilon^{IJK} \tilde{W}^{I \nu}_\mu W^{J \rho}_\nu W^{K \mu}_\rho \ .
\end{align}
\end{linenomath}
The modified couplings and gauge fields introduced in Eqns.\ref{5.16a}, \ref{5.16b} do not cancel the new contribution  from these operators suppressed by $v^2_T/\Lambda^2$ to the $CP$ violating $\tilde{\theta}$ parameters. These extra contributions strongly indicate
that without fine tuning the Wilson coefficients $C_{H\tilde{G}},C_{H\tilde{W}},C_{H \tilde{B}},C_{H \tilde{W} B}$ are suppressed by a large $CP$ violating scale and can be neglected, similar to the treatment of $\mathcal{L}_5$.

\subsubsection{$h \to WW$ and $h \to ZZ$}
The relevant $CP$-even Lagrangian terms are
\begin{linenomath}
\begin{align}
\mathcal{L} &= (D_\mu H)^\dagger (D^\mu H) - \frac{1}{4} \left(W^I_{\mu\nu} W^{I \mu\nu} + B_{\mu\nu} B^{\mu\nu} \right), \nn
&\hspace{0.4cm} + C_{HW} \, Q_{HW} + C_{HB} \, Q_{HB} + C_{HWB} Q_{HWB} + C_{HD} \, Q_{HD},
\label{gaugebosontree}
\end{align}
\end{linenomath}
which lead to the interactions
\begin{linenomath}
\begin{align}
\mathcal{L}&=\frac12 \gcw^2 v_T     h \, W^+_\mu \, W^-_\mu \, \left[1 + \ckin  \right] +  C_{HW}  v_T  h  \, W^+_{\mu \, \nu} \, W^-_{\mu \, \nu}.
\label{5.41}
\end{align}
\end{linenomath}
for the $W$, and
\begin{linenomath}
\begin{align}
\mathcal{L}&=\frac14 (\gcw^2+\gcb^2)v_T h (\mathcal{Z}_\mu)^2 \left[1 + \ckin +v_T^2 C_{HD} \right] +\frac12  \gcb \gcw v_T^3 h (\mathcal{Z}_\mu)^2 C_{HWB} \nn
& +  v_T  h (\mathcal{Z}_{\mu\nu})^2 \left( \frac{\gcw^2 C_{HW}+ \gcb^2 C_{HB} +\gcb \gcw C_{HWB}}{\gcw^2+\gcb^2} \right)
\label{5.42}
\end{align}
\end{linenomath}
for the $Z$. Normalizing the SM $\tilde{\theta}$ operators by two powers of the appropriate gauge coupling so that Eqns.\ref{5.16a}, \ref{5.16b} do not introduce extra terms, the $\cancel{CP}$ contributions are
\begin{linenomath}
\begin{align}
\mathcal{L}^{\cancel{CP}}_6 =  C_{H\tilde{W}}  v_T  h  \, \tilde{W}^+_{\mu \, \nu} \, W^-_{\mu \, \nu} + v_T  h (\mathcal{\tilde{Z}}_{\mu\nu} \, \mathcal{Z}^{\mu\nu}) \left( \frac{\gcw^2 C_{H\tilde{W}}+ \gcb^2 C_{H \tilde{B}} +\gcb \gcw C_{H\tilde{W}B}}{\gcw^2+\gcb^2} \right).
\end{align}
\end{linenomath}

\subsubsection{$h \to \gamma \, \gamma$, $h \to \gamma \,Z$ and $h \rightarrow gg$}

The CP even and odd couplings of $h\rightarrow \gamma \, \gamma$ and $h\rightarrow \gamma \,Z$ are given by \cite{Manohar:2006gz}
\begin{linenomath}
\begin{align}
\mathcal{L}&= h \, v_T \, \bar{e}^2 \, \left[C_{\gamma \, \gamma} \, \mathcal{A}^{\mu \, \nu} \, \mathcal{A}_{\mu \, \nu} + \tilde{C}_{\gamma \, \gamma} \, \tilde{\mathcal{A}}^{\mu \, \nu} \, \mathcal{A}_{\mu \, \nu}
+ C_{\gamma \, Z} \, \mathcal{A}^{\mu \, \nu} \, \mathcal{Z}_{\mu \, \nu} + \tilde{C}_{\gamma \, Z} \, \tilde{\mathcal{A}}^{\mu \, \nu} \, \mathcal{Z}_{\mu \, \nu} \right].
\end{align}
\end{linenomath}
Here 
\begin{linenomath}
\begin{align}
C_{\gamma \, \gamma} &= \frac{C_{HW}}{\bar{g}^2_2} + \frac{C_{HB}}{\bar{g}^2_1} - \frac{C_{HWB}}{\bar{g}_1 \, \bar{g}_2}, \\
\tilde{C}_{\gamma \, \gamma} &= \frac{C_{H\tilde{W}}}{\bar{g}^2_2} + \frac{C_{H\tilde{B}}}{\bar{g}^2_1} - \frac{C_{H\tilde{W}B}}{\bar{g}_1 \, \bar{g}_2}, \\
C_{\gamma \, Z} &= \frac{1}{\bar{g}_1 \, \bar{g}_2} \left( C_{HW} - C_{HB}\right) - \left(\frac{1}{2 \,\bar{g}_1^2} - \frac{1}{2 \,\bar{g}_2^2} \right) C_{HWB} , \\
\tilde{C}_{\gamma \, Z} &=\frac{1}{\bar{g}_1 \, \bar{g}_2} \left( C_{H\tilde{W}} - C_{H\tilde{B}}\right) - \left(\frac{1}{2 \,\bar{g}_1^2} - \frac{1}{2 \,\bar{g}_2^2} \right) C_{H\tilde{W}B}. 
\end{align}
\end{linenomath}
The CP even and odd couplings of $h\rightarrow gg$ are trivially
\begin{linenomath}
\begin{align}
\mathcal{L} =  h \, v_T \, \left[C_{HG} \, \mathcal{G}^{\mu \, \nu} \, \mathcal{G}_{\mu \, \nu} + C_{H\tilde{G}} \,  \mathcal{\tilde{G}}^{\mu \, \nu} \, \mathcal{G}_{\mu \, \nu} \right].
\end{align}
\end{linenomath}
\subsubsection{Dipoles and Higgs dipole interactions}

In the broken phase the dipole interactions with neutral gauge bosons are
\begin{linenomath}
\begin{align}
\mathcal{L} &=\frac{\bar{e} \, (v+ h)}{\sqrt 2} \mathcal{C}_{\substack{\psi \gamma \\ rs }}\  \overline \psi_{r} \sigma^{\mu \nu} P_R \psi_{s}\, \mathcal{A}_{\mu \nu} 
+ \frac{\bar{e} \, (v+ h)}{\sqrt 2} \mathcal{C}_{\substack{\psi \mathcal{Z} \\ rs }}\  \overline \psi_{r} \sigma^{\mu \nu} P_R \psi_{s}\, \mathcal{Z}_{\mu \nu}, \nn
&+ \frac{(v+ h) }{\sqrt 2} C_{\substack{d G \\ rs }}\  \overline d_{r} \sigma^{\mu \nu} T_A \, P_R d_{s}\, \mathcal{G}^A_{\mu \nu}
+ \frac{ (v+ h) }{\sqrt 2} C_{\substack{u G \\ rs }}\  \overline u_{r} \sigma^{\mu \nu} T_A P_R u_{s}\, \mathcal{G}^A_{\mu \nu} + h.c.
\end{align}
\end{linenomath}
where $r$ and $s$ are flavor indices  and $\psi = \{e, u, d\}$ so that
\begin{linenomath}
\begin{align}
\mathcal{C}_{\substack{e \gamma \\ rs }} &= \left[\mathcal{U}(e,L)^\dagger\left( \frac{C_{\substack{e B}}}{g_1}  - \frac{C_{\substack{e W  }}}{g_2}\right) \mathcal{U}(e,R)\right]_{rs}&
\mathcal{C}_{\substack{e Z \\ rs }} &= - \left[\mathcal{U}(e,L)^\dagger \left(\frac{C_{\substack{e B}}}{g_2} + \frac{C_{\substack{e W}}}{g_1}\right) \mathcal{U}(e,R) \right]_{rs}  \nn
\mathcal{C}_{\substack{d \gamma \\ rs }} &=  \left[\mathcal{U}(d,L)^\dagger\left( \frac{C_{\substack{d B}}}{g_1}  - \frac{C_{\substack{d W  }}}{g_2}\right) \mathcal{U}(d,R)\right]_{rs}&
\mathcal{C}_{\substack{d Z \\ rs }} &=- \left[\mathcal{U}(d,L)^\dagger \left(\frac{C_{\substack{d B}}}{g_2} + \frac{C_{\substack{d W}}}{g_1}\right) \mathcal{U}(d,R) \right]_{rs}  \nn
\mathcal{C}_{\substack{u \gamma \\ rs }} &=   \left[\mathcal{U}(u,L)^\dagger\left( \frac{C_{\substack{u B}}}{g_1}  + \frac{C_{\substack{u W  }}}{g_2}\right) \mathcal{U}(u,R)\right]_{rs}&
\mathcal{C}_{\substack{u Z \\ rs }} &= - \left[\mathcal{U}(u,L)^\dagger \left(\frac{C_{\substack{u B}}}{g_2} - \frac{C_{\substack{u W}}}{g_1}\right) \mathcal{U}(u,R) \right]_{rs}.
\label{2.9}
\end{align}
\end{linenomath}
$C_{uW}$ has the opposite sign for $u$-type quarks in Eq.~(\ref{2.9}) because of the opposite sign for $T_{3L}$. Note that $\sigma_{\mu \,\nu} = i \, \left[\gamma_\mu, \gamma_\nu \right]/2$.
The dipole interactions with charged gauge bosons are
\begin{linenomath}
\begin{align}
\mathcal{L} &= (v+ h) \, \left[ \overline \nu_r \, \sigma^{\mu \nu} \, P_R \, e_s \, \mathcal{W}^+_{\mu \nu} \mathcal{C}_{\substack{e \, W \\ rs }}
+ \overline u_r \, \sigma^{\mu \nu} \, P_R \, d_s \, \mathcal{W}^+_{\mu \nu} \, \mathcal{C}_{\substack{d \, W \\ rs}} 
+ \overline d_r \, \sigma^{\mu \nu} \, P_R \, u_{s}\, \mathcal{W}^-_{\mu \nu} \, \mathcal{C}_{\substack{u \, W \\ rs}} \right]+ h.c.
\end{align}
\end{linenomath}
where
\begin{linenomath}
\begin{align}
\mathcal{C}_{\substack{e \, W \\ rs }} &= \left[\mathcal{U}(\nu,L)^\dagger \, C_{\substack{e \, W}} \, \mathcal{U}(e,R)\right]_{rs}& \mathcal{C}_{\substack{d \, W \\ rs}} =  \left[\mathcal{U}(u,L)^\dagger \, C_{\substack{d \, W}} \, \mathcal{U}(d,R)\right]_{rs} \nn
\mathcal{C}_{\substack{u \, W \\ rs}}  &= \left[\mathcal{U}(d,L)^\dagger \, \mathcal{C}_{\substack{u \, W}} \, \mathcal{U}(u,R)\right]_{rs}.
\end{align}
\end{linenomath}
\subsubsection{$(h + v)^2\, V \bar{\psi} \, \psi$ interactions}

In the broken phase the interactions of the Higgs with  fermions and an associated gauge boson are given by
\begin{linenomath}
\begin{align}
\mathcal{L} &= \frac{\sqrt{\bar{g}_1^2+ \bar{g}_2^2}}{2} \, (h + v_T)^2 \, \mathcal{Z}_\mu \, \bar{\nu}_{r} \, \gamma^\mu P_L \nu_s  \left[\mathcal{U}(\nu,L)^\dagger \, \left(C_{\substack{H \ell }}^{(1)} - C_{\substack{H \ell}}^{(3)} \right) \mathcal{U}(\nu,L)\right]_{rs}, \nn
&+ \frac{\sqrt{\bar{g}_1^2+ \bar{g}_2^2}}{2} \, (h + v_T)^2 \, \mathcal{Z}_\mu \, \bar{e}_{r} \, \gamma^\mu P_L e_s  \left[\mathcal{U}(e,L)^\dagger \, \left(C_{\substack{H \ell }}^{(1)} + C_{\substack{H \ell}}^{(3)} \right) \mathcal{U}(e,L)\right]_{rs}, \nn 
&+\frac{\sqrt{\bar{g}_1^2+ \bar{g}_2^2}}{2} \, (h + v_T)^2 \, \mathcal{Z}_\mu \, \bar{u}_{r} \, \gamma^\mu P_L u_s  \left[\mathcal{U}(u,L)^\dagger \, \left(C_{\substack{H q }}^{(1)} - C_{\substack{H q}}^{(3)} \right) \mathcal{U}(u,L)\right]_{rs}, \nn
&+ \frac{\sqrt{\bar{g}_1^2+ \bar{g}_2^2}}{2} \, (h + v_T)^2 \, \mathcal{Z}_\mu \, \bar{d}_{r} \, \gamma^\mu P_L d_s  \left[\mathcal{U}(d,L)^\dagger \, \left(C_{\substack{H q }}^{(1)} + C_{\substack{H q}}^{(3)} \right) \mathcal{U}(d,L)\right]_{rs}, \nn 
& + \frac{\sqrt{\bar{g}_1^2+ \bar{g}_2^2}}{2} \, (h + v_T)^2 \, \mathcal{Z}_\mu \, \bar{\psi}_{r} \, \gamma^\mu P_R \, \psi_s \,  \left[\mathcal{U}(\psi,R)^\dagger \, C_{\substack{H \psi}} \, \mathcal{U}(\psi,R)\right]_{rs}, \nn
&- \frac{\bar{g}_2}{\sqrt{2}}  \, (h + v_T)^2 \,  \mathcal{W}^+_\mu \, \bar{\nu}_r \, \gamma^\mu P_L  \,  e_s \,\left[\mathcal{U}(\nu,L)^\dagger \, C_{\substack{H \ell}}^{(3)} \, \mathcal{U}(e,L)\right]_{rs} \nn
&- \frac{\bar{g}_2}{\sqrt{2}}  \, (h + v_T)^2 \,  \mathcal{W}^+_\mu \, \bar{u}_r \, \gamma^\mu P_L  \,  d_s \,\left[\mathcal{U}(u,L)^\dagger \, C_{\substack{H q}}^{(3)} \, \mathcal{U}(d,L)\right]_{rs},\nn
&+ \frac{i \, \bar{g}_2}{2}  \, (h + v_T)^2 \,  \mathcal{W}^+_\mu \, \bar{u}_r \, \gamma^\mu P_R  \,  d_s \,\left[\mathcal{U}(u,R)^\dagger \, C_{\substack{H ud}} \, \mathcal{U}(d,R)\right]_{rs} + h.c
\end{align}
\end{linenomath}
where $\psi = \{u,d,e \}$.
\subsubsection{TGC parameters}
The off-shell Triple gauge coupling parameters are given by
\begin{linenomath}
\begin{align}
\left(-\mathcal{L}_{TGC} \right)/ \bar{g}_{VWW} &=i \bar{g}_{1}^{V} \left( \mathcal{W}_{\mu \nu}^{+} \mathcal{W}^{- \mu} - \mathcal{W}_{\mu \nu}^{-} \mathcal{W}^{+ \mu}\right)\mathcal{V}^{\nu} +i \bar{\kappa}_{V} \mathcal{W}^{+}_{\mu}\mathcal{W}^-_{\nu}\mathcal{V}^{\mu \nu}, \\
&+ i \frac{\bar{\lambda}_{V}}{\bar{M}^2_W}\mathcal{V}^{\mu \nu} \mathcal{W}^{+ \rho}_{\nu}\mathcal{W}^{-}_{\rho \mu} \nonumber
\end{align}
\end{linenomath}

where $V = \{\mathcal{Z},\mathcal{A} \}$. In the SM $g_{\mathcal{AWW}} = e$ and $g_{\mathcal{ZWW}} = g_2 \, c_{\theta}$.
In the SMEFT the canonically normalized couplings are modified to $\bar{g}_{\mathcal{AWW}} = \bar{e}$ and $\bar{g}_{\mathcal{ZWW}} = \bar{g}_2 \, \bar{c}_{\theta}$
and the shifts compared to these normalized couplings are
\begin{linenomath}
\begin{align}
\delta \bar{g}_1^{\mathcal{A}}  &=- \delta \bar{\kappa}_{\mathcal{A}} = - \frac{v_T^2}{2}\frac{c_{\bar{\theta}}}{s_{\bar{\theta}}} C_{HWB},  \quad \quad \quad &\delta \bar{g}_1^{\mathcal{Z}} = -\delta \kappa_{\mathcal{Z}} = \frac{v_T^2}{2}\frac{s_{\bar{\theta}}}{c_{\bar{\theta}}} C_{HWB}, 
\end{align}
\end{linenomath}
and 
\begin{linenomath}
\begin{align}
\delta \bar{\lambda}_{\mathcal{A}} &=  6 \, s_{\bar{\theta}} \, C_W \, \frac{\bar{M}^2_W}{\bar{g}_{\mathcal{AWW}}},  \quad \quad \quad &\delta \bar{\lambda}_{\mathcal{Z}} =  6 \, c_{\bar{\theta}} \, C_W \frac{\bar{M}^2_W}{\bar{g}_{\mathcal{ZWW}}}.
\end{align}
\end{linenomath}
An important check of gauge invariance in TGC shifts is that the relationships
\begin{linenomath}
\bea
\bar{\kappa}_{\mathcal{Z}} = \bar{g}_1^{\mathcal{Z}} - (\bar{\kappa}_{\mathcal{A}}-1) \, t_{\bar{\theta}}^2, \quad \quad \bar{\lambda}_{\mathcal{Z}} =  \bar{\lambda}_{\mathcal{A}},
\eea
\end{linenomath}
are respected when the shifts in the Lagrangian parameters are expressed in terms of the SM parameters.
These shifts respect these relationships. 

\subsection{Summary of mass eigenstate interactions and symmetries}

$\mathcal{L}_6$ has $2499$ parameters in general \cite{Alonso:2013hga}. Clearly restricting to
a Minimal Flavour Violating (MFV) scenario \cite{Chivukula:1987py,D'Ambrosio:2002ex,Cirigliano:2005ck}, which imposes a $\rm U(3)^5$  flavour symmetry broken only by the SM Yukawas
is desirable. This reduces the number of parameters to $76$.
Assuming that $\rm CP$ violating effects can also be neglected, the number of parameters is restricted to $53$ for $\mathcal{L}_6$ \cite{Alonso:2013hga}. This is a reasonable symmetry based 
limit to assume.\footnote{Custodial symmetry is broken by gauge interactions in the SM and the mass splitting of fermion doublet fields.
The number of parameters removed due to this strongly broken symmetry being assumed are negligible compared to the effects of the $\rm CP$ even and MFV assumptions.}
In this symmetric case $\left[ {\cal Y}_\psi \right]_{rs} \in \mathbb{R}$ and
\begin{linenomath}
\bea
\left[ {\cal Y}_\psi \right]_{rs} =  \delta_{rs} \, \frac{\hat{m}^r_\psi}{v_T} \,  \left[ 1+ \ckin   - v^2\,  \mathcal{C}_{\substack{\psi H}} \right], \quad {\rm where} \quad \mathcal{C}_{\substack{\psi H}} [Y_\psi]_{rs} =  {\rm Re} \left[C^*_{\substack{\psi H \\ rs}}\right].
\eea
\end{linenomath}
Further all Wilson coefficients for operators with dual fields (denoted with tilde subscripts or superscripts) are neglected. The dipole and Higgs dipole interactions are all flavour diagonal proportional to the corresponding
fermion mass and real. The $(h + v)^2\, \mathcal{Z} \bar{\psi} \, \psi$ interactions are flavour diagonal while the $(h + v)^2\, \mathcal{W}^+ \bar{\psi} \, \psi$ interactions are proportional to $V_{CKM}$ or $\mathcal{U}_{PMNS}^\dagger$. Finally, flavour violating interactions in the Class 8 operators follow an MFV pattern \cite{Chivukula:1987py,D'Ambrosio:2002ex,Cirigliano:2005ck}.

These symmetries, if assumed in $\mathcal{L}_6$, are broken at least by the SM interactions, which violate these symmetries. NLO calculations are required to define the perturbative breaking of these symmetric limits. 

\subsection{Input parameters and defining conditions}

At variance with what is done in alternate LO constructions, we have not imposed the following conditions on the mass eigenstate construction (including $\mathcal{L}_6$ corrections):
\begin{itemize}
\item{Tree-level relations between the electroweak parameters and a choice of input parameter set (IPS) are the same as the SM ones.}
\item{Two-derivative self-interactions of the Higgs boson are absent.}
\item{For each fermion pair, the coefficients of the $h V \bar{\psi} \, \psi$, $h^2 V \bar{\psi} \, \psi$ interaction terms are equal to the vertex correction of $V  \bar{\psi} \, \psi$.}
\end{itemize}
These conditions are not required to interpret the data in the SMEFT. For example the two-derivative self-interactions of the Higgs boson is a trivial contact
interaction that can be directly included in an analysis. \xxxr{It was proven in Ref.\cite{Burgess:2010zq} that no {\it gauge independent manipulations and field redefinitions} allow the second condition to be 
imposed. When conditions such as these are imposed in a manner that does not respect the equivalence theorem, this requires a loss of generality. }As a consequence, these conditions cannot be imposed without introducing technical complications in a LO approach that make NLO calculations harder to develop, or simply impossible, in our view.

Considering condition one above, we emphasize that an operator basis is IPS independent.
If one were to modify the construction of $\mathcal{L}_{SMEFT}$ to make some relationships
to a particular IPS the same in the SM and the SMEFT with algebraic manipulations that were only defined classically (i.e. at LO), this would make such a construction
an example of a ``phenomenological effective Lagrangian" that was then limited to LO. If the algebraic manipulations were not the same 
for all possible IPS sets that could be chosen this would tie such a construction to a specific IPS.  
Claims of intuitive connections to LHC Higgs and EWPD observables in such approaches should be considered with great care and skepticism. 
We advocate here to not attempt to intrinsically tie a phenomenological Lagrangian construction to any specific IPS, for a series of reasons:
\begin{itemize}
\item{Monte Carlo programs do not all use the same IPS. Further, the IPS $\{\alpha_{ew}, G_F, M_Z \}$ is not in common use when
automated calculations for the SM beyond LO are generated to define the SM event rate in a measurement.
Before any SMEFT implementation is used it must first be checked what IPS set or sets are used to define the SM event rate in the measurement of interest.
If a construction tied to the specific IPS $\{\alpha_{ew}, G_F, M_Z \}$ were to be used it must be confirmed that all simulation tools and SM results only use this specific IPS or inconsistent
results will be reported.} 
\item{When  the IPS $\{\alpha_{ew}, G_F, M_Z \}$ is used in the analysis of ``high'' energy data it is afflicted with
hadronic uncertainties entering already at the one loop level and arising because it must be
``run up'' from low energy, crossing the hadronic resonance region.
The Fermi coupling constant, obtained from the muon lifetime, does not suffer from this
disadvantage (even in the full SM one loop hadronic effects are mass 
suppressed)~\cite{vanRitbergen:1999fi}.}
\end{itemize}

The parameters $v_T$, $\bar{g}_1$, $\bar{g}_2$, $\bar{e}$, $s_{\bar{\theta}}$, $c_{\bar{\theta}}$ etc. in the Lagrangian do have to be assigned numerical values consistent with some IPS.
This is sometimes known as a ``finite renormalization". This is distinct from rotating to the mass eigenstate fields in the canonically normalized SMEFT and does not require the conditions above be imposed
in a gauge dependent manner.
We now illustrate how a straightforward LO implementation is related to the IPS $\{\alpha_{ew}, G_F, M_Z \}$ in the $\rm U(3)^5$ limit for tree level gauge boson fermion interactions.
\subsubsection{Input parameters measurements}
Define the local effective interaction for muon decay as
\begin{linenomath}
\begin{align}
\mathcal{L}_{G_F} =  -\frac{4G_{F}}{\sqrt{2}} \, \left(\bar{\nu}_\mu \, \gamma^\mu P_L \mu \right) \left(\bar{e} \, \gamma_\mu P_L \nu_e\right).
\end{align}
\end{linenomath}
$G_F$ is defined as the following parameter measured in $\mu$ decay, $\mu^- \rightarrow e^- + \bar{\nu}_e + \nu_\mu$. 
In the SMEFT ($e$ and $\mu$ are generation indices 1 and 2, and are not summed over)
\begin{linenomath}
\begin{align}
-\frac{4G_{F}}{\sqrt{2}} &=  -\frac{2}{v_T^2} +  \left(C_{\substack{ll \\ \mu ee \mu}} +  C_{\substack{ll \\ e \mu\mu e}}\right) - 2 \left(C^{(3)}_{\substack{Hl \\ ee }} +  C^{(3)}_{\substack{Hl \\ \mu\mu }}\right).
\label{gfermi}
\end{align}
\end{linenomath}
The parameter $\alpha_{ew}$ is measured in the Thompson ($p^2 \rightarrow 0$) limit and discussed in Section \ref{alphaew}, and $M_Z$ is defined in the resonance pole scan of LEP measurements.

\subsubsection{Gauge boson couplings for the $\alpha$ IPS}
Our notational conventions are that shifts due to the
SMEFT are denoted as $\delta X = (X)_{SMEFT} - X_{SM}$ for a parameter $X$.\footnote{See Refs.~\cite{Grinstein:1991cd,Han:2004az,Burgess:1993vc,Bamert:1996px,Burgess:1993qk,Pomarol:2013zra} for the development of this approach and Refs. \cite{Berthier:2015gja,Berthier:2015oma} for details.}
Measured input observables or parameters directly defined by combinations of input observables are denoted with hat superscripts.
The shifts in the commonly appearing Lagrangian parameters $M_Z$, $M_W$, $G_F$, $s_\theta^2$ are 
\begin{linenomath}
\bea
\delta M_Z^2 &\equiv&  \frac{1}{2 \, \sqrt{2}} \, \frac{\hat{m}_Z^2}{\hat{G}_F} C_{HD} + \frac{2^{1/4} \sqrt{\pi} \, \sqrt{\hat{\alpha}} \, \hat{m}_Z}{\hat{G}_F^{3/2}} C_{HWB}, \\
\delta M_W^2 &=& -\hat{m}_W^2 \left(\frac{\delta s_{\that}^2}{s_{\that}^2}+\frac{c_{\that}}{s_{\that} \sqrt{2} \hat{G}_F}C_{HWB} + \sqrt{2} \delta G_F\right),\\
\delta G_F &=&  \frac{1}{\sqrt{2} \,  \hat{G}_F} \left(\sqrt{2} \, C^{(3)}_{\substack{Hl}} - \frac{C_{\substack{ll}}}{\sqrt{2}}\right), \\
\delta s_\theta^2 &=&  - \frac{s_\that \, c_\that}{2 \, \sqrt{2} \, \hat{G}_F (1 - 2 s^2_\that)} \left[s_\that \, c_\that \, (C_{HD} + 4 \, C^{(3)}_{\substack{H l}} - 2 \, C_{\substack{ll}}) 
+ 2 \, C_{HWB} \right].
\eea
\end{linenomath}
These shifts lead to modifications of the $Z$ couplings with the normalization
\begin{linenomath}
\bea
\mathcal{L}_{Z,eff}  =  g_{Z,eff}  \,   \left(J_\mu^{Z \ell} Z^\mu + J_\mu^{Z \nu} Z^\mu + J_\mu^{Z u} Z^\mu +  J_\mu^{Z d} Z^\mu \right),
\eea 
\end{linenomath}
where $g_{Z,eff} = - \, 2 \, 2^{1/4} \, \sqrt{\hat{G}_F} \, \hat{m}_Z$, $(J_\mu^{Z x})^{pr} = \bar{x}_p \, \gamma_\mu \left[(\bar{g}^{x}_V)_{eff}^{pr}- (\bar{g}^{x}_A)_{eff}^{pr} \, \gamma_5 \right] x_r$ for $x = \{u,d,\ell,\nu \}$.
In general, these currents are matrices in flavour space. When we restrict our attention to the case of a MFV scenario $(J_\mu^{Z x})_{pr} \simeq (J_\mu^{Z x}) \delta_{pr}$.
In the Warsaw basis, the effective axial and vector couplings are modified from the SM values by a shift 
\begin{linenomath}
\bea
\delta (g^{x}_{V,A})_{pr} = (\bar{g}^{x}_{V,A})^{eff}_{pr} - (g^{x}_{V,A})^{SM}_{pr}, 
\eea
\end{linenomath}
where
\begin{linenomath}
\bea\label{higherdgvga}
\delta (g^{\ell}_V)_{pr}&=&\delta \bar{g}_Z \, (g^{\ell}_{V})^{SM}_{pr} - \frac{1}{4 \sqrt{2} \hat{G}_F} \left(C_{\substack{H e \\pr}} + C_{\substack{H l \\ pr}}^{(1)} + C_{\substack{H l \\ pr}}^{(3)} \right) - \delta s_\theta^2, \\
\delta(g^{\ell}_A)_{pr}&=&\delta \bar{g}_Z \, (g^{\ell}_{A})^{SM}_{pr} + \frac{1}{4 \, \sqrt{2} \, \hat{G}_F} 
\left( C_{\substack{H e \\pr}} - C_{\substack{H l \\ pr}}^{(1)} - C_{\substack{H l \\ pr}}^{(3)} \right),  \\
\delta (g^{\nu}_V)_{pr}&=&\delta \bar{g}_Z \, (g^{\nu}_{V})^{SM}_{pr} - \frac{1}{4 \, \sqrt{2} \, \hat{G}_F} \left( C_{\substack{H l \\ pr}}^{(1)} - C_{\substack{H l \\ pr}}^{(3)} \right), 
\\
\delta(g^{\nu}_A)_{pr}&=&\delta \bar{g}_Z \,(g^{\nu}_{A})^{SM}_{pr}  - \frac{1}{4 \, \sqrt{2} \, \hat{G}_F} 
\left(C_{\substack{H l \\ pr}}^{(1)} - C_{\substack{H l \\ pr}}^{(3)} \right),  \\
\delta (g^{u}_V)_{pr}&=&\delta \bar{g}_Z \, (g^{u}_{V})^{SM}_{pr}  + 
\frac{1}{4 \, \sqrt{2} \, \hat{G}_F} \left(- C_{\substack{H q \\ pr}}^{(1)} + \, C_{\substack{H q \\ pr}}^{(3)} -C_{\substack{H u \\ pr}} \right) + \frac{2}{3} \delta s_\theta^2,\\
\delta(g^{u}_A)_{pr}&=&\delta \bar{g}_Z \, (g^{u}_{A})^{SM}_{pr}
-\frac{1}{4 \, \sqrt{2} \, \hat{G}_F} \left( C_{\substack{H q \\ pr}}^{(1)} -  \, C_{\substack{H q \\ pr}}^{(3)} - C_{\substack{H u \\ pr}} \right), \\ 
\delta (g^{d}_V)_{pr}&=&\delta \bar{g}_Z \,(g^{d}_{V})^{SM}_{pr} 
-\frac{1}{4 \, \sqrt{2} \, \hat{G}_F} \left( C_{\substack{H q \\ pr}}^{(1)}  +  \, C_{\substack{H q \\ pr}}^{(3)} + C_{\substack{H d \\ pr}} \right) -  \frac{1}{3} \delta s_\theta^2, \\
\delta(g^{d}_A)_{pr}&=&\delta \bar{g}_Z \,(g^{d}_{A})^{SM}_{pr} 
+\frac{1}{4 \, \sqrt{2} \, \hat{G}_F} \left(-C_{\substack{H q \\ pr}}^{(1)}  -  \, C_{\substack{H q \\ pr}}^{(3)} + C_{\substack{H d \\ pr}} \right),
\eea
\end{linenomath}
where
\begin{linenomath}
\bea
\delta \bar{g}_Z =- \frac{\delta G_F}{\sqrt{2}} - \frac{\delta M_Z^2}{2\hat{m}_Z^2} + \frac{s_{\hat{\theta}} \, c_{\hat{\theta}}}{\sqrt{2} \hat{G}_F} \, C_{HWB},
\eea
\end{linenomath}
and similarly the $W$ couplings are defined as
\begin{linenomath}
\bea
\delta(g^{W_{\pm},\ell}_V)_{rr} = \delta(g^{W_{\pm},\ell}_A)_{rr}  &=&  \frac{1}{2\sqrt{2} \hat{G}_F} \left(C^{(3)}_{\substack{H l \\ rr}} + \frac{1}{2} \frac{c_{\hat{\theta}}}{ s_{\hat{\theta}}} \, C_{HWB} \right)
+ \frac{1}{4} \frac{\delta s_\theta^2}{s^2_{\hat{\theta}}},
\eea
\end{linenomath}
\begin{linenomath}
\bea
\delta(g^{W_{\pm},q}_V)_{rr} = \delta(g^{W_{\pm},q}_A)_{rr}  &=&  \frac{1}{2\sqrt{2} \hat{G}_F} \left(C^{(3)}_{\substack{H q \\ rr}} + \frac{1}{2} \frac{c_{\hat{\theta}}}{ s_{\hat{\theta}}} \, C_{HWB} \right)
+ \frac{1}{4} \frac{\delta s_\theta^2}{s^2_{\hat{\theta}}}.
\eea
\end{linenomath}
Here our chosen normalization is $(g^{x}_{V})^{SM} = T_3/2 - Q^x \, s_\theta^2, (g^{x}_{A})^{SM} = T_3/2$ where $T_3 = 1/2$ for $u_i,\nu_i$ and $T_3 = -1/2$ for $d_i,\ell_i$
and $Q^x = \{-1,2/3,-1/3 \}$ for $x = \{\ell,u,d\}$.  
The set of $\delta X$ parameters are not an operator basis for the SMEFT. 


\subsection{Fitting at LO and NLO: constraints and covariance}

The mapping of an experimental constraint to the underlying $C_i$ is based on a linear expansion of a cross section or a pseudo-observable based decomposition of a cross section. 
A fit at LO to mass eigenstate parameters should include a theoretical covariance matrix 
and a theoretical error due to neglected higher order effects in the SMEFT \cite{Berthier:2015gja,Berthier:2015oma,David:2015waa}. 
A fit in terms of the underlying weak eigenstate Wilson coefficients is straightforward and will in general have a much simpler theoretical covariance matrix.

\subsubsection{Digression on theoretical uncertainty}
In the SM, when a particular process is calculated, a common practice is that
a theoretical error is assigned. For example, for parametric and theoretical uncertainties within the SM, see Tab.~1 of
\Bref{Heinemeyer:2013tqa}.  It can be subtle to assign such an error~\cite{David:2013gaa} 
due to the neglect of missing higher order perturbative terms in the SM. 
The need to include theoretical errors when perturbatively expanding the SMEFT is tied to the fact that different truncations of such expansions can
be constructed. Suppose that a given quantity $\mathrm{Q}(a)$ is given in perturbation theory by the following expansion:
\begin{linenomath}
\bq
\mathrm{Q} = a + g\,\Bigl[ a^2 + f_1(a) \Bigr] + g^2\,\Bigl[ a^3 + f_2(a) \Bigr] + \mcO(g^3) 
= {\bar a} + g\,f_1(a) + \mcO(g^2),
\eq
\end{linenomath}
where ${\bar a} = a/(1 - g a)$. Suppose that only the $f_1$ term is actually known. It could be 
decided that ${\bar a}$ is the effective expansion parameter (or that in the full expression we 
change variable $a \to {\bar a}$). This is equivalent, in the truncated expansion, to introducing
\begin{linenomath}
\bq
\mathrm{Q} = \bar{a} + g\,f_1(a) = {\bar a} + g\,f_1({\bar a}),
\eq
\end{linenomath}
which gives $\Delta \mathrm{Q} = g^2\,f'_1(a)$, the difference in the two results due to neglected higher order terms is an estimate of the associated theoretical uncertainty. 
A fit to observables defined in a perturbative expansion must always include an estimate of the missing higher order terms~\cite{Bardin:1999gt}, which specifies
a theoretical uncertainty. 

\subsubsection{The importance of NLO results for theoretical uncertainty}

An excellent example of the importance of theory errors is provided by another effective field theory, NRQED, as discussed in
Refs.\cite{Bauer:2004ve,Manohar:2003xv,Caswell:1985ui,Bodwin:1994jh,Luke:1999kz,Pineda:1998kn,Grinstein:1997gv}.
The Hydrogen hyperfine splitting is measured to fourteen digits, but only computed to seven digits. 
This introduces a theoretical error when using this measurement. Comparatively, the Positronium hyperfine splitting is measured
and computed to eight digits. It would simply be a mistake to give the $H$ hyperfine splitting a weight $10^6$ larger than the $P_s$ hyperfine
splitting in a global fit to the fundamental constants, and to totally ignore theory errors. A careful consideration of NLO effects can help in
avoiding similar errors when using the SMEFT formalism.

Neglecting such considerations has already led to incorrect conclusions. For example, it has been shown that claims of general model independent bounds
at the per-mille level due to the LEP experiments projected into the SMEFT (common in some recent literature) do not hold when considering SMEFT theoretical errors \cite{Berthier:2015gja,Berthier:2015oma}.
This should be unsurprizing, as in EWPD the modifications of the $\PW$ mass, the $\uprho$ parameter and the effective weak-mixing angle are 
loop-induced quantities and a study of their SM deviations requires an analysis at NLO in the SMEFT.
As a result of these developments, constraints on parameters in the SMEFT (that are not symmetry based) are not robustly below LHC sensitivity.

For this reason, it is not advisable to set parameters that contribute to EWPD to zero in LHC analyses in the SMEFT.
The experimental bound should be imposed on these parameters, with a clearly specified theory error.
As a rule of thumb when experimental bounds descend below the $10 \%$ level SMEFT theory errors should not be neglected in an EFT interpretation of the data.

\subsubsection{Covariance due to operator basis in $\mathcal{L}_6$}\label{variancesection}

Consider two mass eigenstate interaction shifts $\delta X_1, \delta X_2$ that contribute to a particular cross section that reports an experimental bound.
Several SMEFT Wilson coefficients  generally contribute to any one observable through $\delta X_1, \delta X_2$.  All such parameters must be retained unless symmetries, or knowledge of the UV theory, allows a reduction. One can directly interpret the data at LO in terms of the underlying Wilson coefficients that are present in $\delta X_1, \delta X_2$ and defined in linear
expansions of these parameters, so long
as theoretical errors are carefully accounted for.

Alternatively fit results can be reported in terms of $\delta X_1, \delta X_2$. However in this case it is critical that a theoretical covariance matrix is included.
As the shifts $\delta X_1, \delta X_2$ are linear in the Wilson coefficients, the bi-linearity property of covariance
can be used to obtain the theoretical covariance matrix directly. Schematically the matrix can be build up for  $\delta X_1 = a C_1 + b C_2 + \cdots$ and $\delta X_2 = c C_1+ d C_3 + \cdots$
as follows
\begin{linenomath}
\bea
Cov \left[\delta X_1,  \delta X_2\right] = a \,c \, Cov [C_1,C_1] + a \,d \, Cov [C_1,C_3] + b \,c \, Cov [C_2,C_1] + b \,d \, Cov [C_2,C_3] + \cdots \nonumber
\eea
\end{linenomath}
Assuming that the $C_1,C_2,C_3$ are independent operators $Cov [C_1,C_1] = Var[C_1]$ and all other entries vanish.
The appropriate covariance matrix can be constructed so long as a theoretical error is included for each of the terms in the perturbative expansion of the $\delta X$.
Estimating a theoretical error for these terms to obtain the individual variances requires an estimate of neglected NLO corrections.
A NLO mapping can be carried out in the same manner. The only modification is the use of NLO formuli in the expansion of the cross section and smaller theoretical errors, as we illustrate below. 
Fits to mass eigenstate parameters in general have very non trivial covariance matricies (due to gauge invariance of the underlying operator basis) that have to be defined. The required
theoretical errors can only be estimated by an understanding of NLO corrections to a LO formalism.

\subsubsection{Fitting at LO or NLO?}

A NLO  treatment of the data is always advisable if the required theoretical results are available.
NLO analyses are required to consistently map
lower energy measurements in the SMEFT to the cut off $\mu = \Lambda$, or to consistently 
combine data sets measured at different effective scales ($\mu_1 \neq \mu_2$). 
Whether or not a NLO treatment of the data is {\it required} in the SMEFT is defined by three considerations:

\begin{itemize}
\item{What is the cut off scale ($\Lambda$) and what is the matching pattern of Wilson coefficients into the SMEFT?} 

\item{What is the experimental precision that will be reached in a measurement?}

\item{How will a bound projected into the SMEFT formalism at LO be used?}
\end{itemize}

Considering the first question, it is interesting to consider the cases where
$1 \, {\rm TeV} \lesssim \Lambda/\sqrt{\tilde{C}_i} \lesssim 3 \, {\rm TeV}$.
In these cases, deviations in processes measured at the LHC could possibly be observable.
If deviations are seen then a NLO analysis is well motivated to learn as precisely as possible about
the underlying physics sector through the measured deviation. 
Cut off scales of this form are not implausible or ruled out. On the contrary they are well motivated by the Hierarchy problem. Further model building exercises for decades have indicated 
that such cut off scales are not robustly ruled out when considering EWPD. If the ratio $ \Lambda/\sqrt{\tilde{C}_i} $ lies in this interesting range,
the effect of NLO corrections are clearly not negligible 
\cite{Grojean:2013kd,Berthier:2015gja,Berthier:2015oma,David:2015waa,Englert:2014cva,Hagiwara:1993ck,Hagiwara:1993qt,Alam:1997nk,Passarino:2012cb,Wells:2015cre,Gauld:2015lmb,Hartmann:2015oia,Hartmann:2015aia,Ghezzi:2015vva}. 
Considering the second question, as we have stressed, when experimental precision starts to reach the $10\%$ level a NLO analysis should be pursued.
The answer to the third question differs among analyses and authors but in general NLO results will always be useful to
authors interested in LO results while the converse is not true. 
\subsubsection{Theory errors in a LO formalism on the IPS}\label{alphaew}
As a specific example of a theory error to include in a LO analysis, any LO approach does not take into account that the scales characterizing the measurements of the input parameters $\alpha_{ew}, G_F, M_Z$ differ.
Consider the error 
introduced due to the neglect of this NLO effect in the SMEFT, compared to the errors quoted on  $\alpha_{ew}$ in the SM.
This parameter is measured at low energies in the $p^2 \rightarrow 0$ limit.\footnote{$\alpha_{ew}$  is frequently extracted in the Thompson limit $p^2 \rightarrow 0$ when probing some Coulomb potential
of a charged particle, for example in a measurement of $g-2$ for the electron or muon. Recently, extractions with a competitive error budget have emerged where $\alpha_{ew}$ is extracted
from the measured ratio of $\hbar/M_{atom}$ via the recoil velocity for a stable atom, such as ${\rm Rb}^{87}$ \cite{Hanneke:2008tm} or ${\rm C_s}$ \cite{2002PhST..102...82W}. The important point is to realize that
this input parameter differs in the SM and in the SMEFT at NLO.} The value of this input parameter is given in Table 2.
In the SMEFT, the running of $\alpha_{ew}$ is modified compared to the SM as given in Ref.\cite{Jenkins:2013zja}.
As a simple approximation of the error introduced in the SMEFT, one finds that the neglected NLO SMEFT correction to $\alpha_{ew}$ is then
\begin{linenomath}
\bea
\frac{(\Delta \alpha_{ew})_{SMEFT}}{(\Delta \alpha_{ew})_{SM}} \simeq - 250 \, \left(\frac{1 {\rm TeV}}{\Lambda} \right)^2 \tilde{C}_{HB} - 80 \, \left(\frac{1 {\rm TeV}}{\Lambda} \right)^2 \tilde{C}_{HW},
\eea
\end{linenomath}
running from $p^2 \sim 1 \, {\rm GeV^2}$ to $m_h$.\footnote{This is only an approximation, as formally all of the SM states with masses $m^2 \gg p^2$ should be integrated out in sequence
when running down from the high scale.} 
Here $(\Delta \alpha_{ew})_{SM}$ is the SM error quoted in the Table. Depending on $\tilde{C}_{HB}$ and $\tilde{C}_{HW}$ and $\Lambda$, which are unknown, the neglected
NLO SMEFT effects can lead to an error on this input parameter far larger than in the SM. 
This should be completely unsurprising. Neglected NLO effects in the SMEFT in this case include corrections of order $ g_{1,2}^2 v_T^2/ (16 \, \pi^2) \, \Lambda^2$. 
The theoretical errors due to such neglected effects can 
obviously compete with the SM theoretical errors, introduced in a QED calculation out to {\it tenth order} in the SM.
Similarly, neglected NLO corrections on the other input parameters modify their theoretical error.
\begin{center}
\begin{table}
\centering
\tabcolsep 8pt
\begin{tabular}{|c|c|c|}
\hline
Parameter & Input Value & Ref.  \\ \hline
$\hat{m}_Z$ & $91.1875 \pm 0.0021$ & \cite{Z-Pole,Agashe:2014kda,Mohr:2012tt} \\
$\hat{G}_F$ & $1.1663787(6) \times 10^{-5} $ &  \cite{Agashe:2014kda,Mohr:2012tt} \\
$\hat{\alpha}_{ew}$ & $1/137.035999074(94) $ &  \cite{Hanneke:2008tm,Agashe:2014kda,Mohr:2012tt,2011PhRvL.106h0801B} \\
\hline
\end{tabular}
\caption{Current experimental best estimates of $\alpha_{ew}, G_F, M_Z$. }
\end{table}
\end{center}
\subsubsection{Approximating unknown SMEFT theory errors}
Various ways exist to estimate SMEFT theory errors. One can 
compute the same observable with different ``options'', \eg linearization or 
quadratization of the squared matrix element, resummation or expansion of the (gauge invariant)
fermion part in the wave function factor for the external legs, variation of the renormalization 
scale, $G_{\mathrm{F}}$ renormalization scheme or $\alpha\,$-scheme, \etc 

A conservative estimate of the associated theoretical uncertainty is obtained by taking the
envelope over all ``options''; the interpretation of the envelope is a log-normal distribution
(commonly done in the experimental community) or a flat Bayesian 
prior~\cite{Cacciari:2011ze,David:2013gaa} (a solution preferred in a large part of the theoretical 
community).

To properly characterize the perturbative error, it is essential to calculate at least to one loop 
order in the SMEFT, including the leading insertion of operators in $\mcL_6$. Until such 
calculations are performed, conservative theoretical errors should be
applied to theoretical relations in the SMEFT. Further, the introduction of a ``non-perturbative'' 
error, due to $\mcL_8$ when bounding $\mcL_6$ should be done. 
In Eqn.\ref{tableNLO}, the $g^3\,g^2_6\,\mcA^{(6)}_{3,2,1}$ terms can be used as estimators 
of missing higher order non-perturbative terms in the SMEFT.
This approach is not particularly novel, but is simply the obvious extension of the widely 
accepted approach to assigning theoretical error in the SM to the SMEFT.

As a specific example, a reasonable approximation of a theoretical error to introduce for an observable $i$ when fitting 
to the leading parameters in $\mcL_6$, is given by~\cite{Berthier:2015gja,Berthier:2015oma}
\begin{linenomath}
\bea
\Delta_{SMEFT}^i(\Lambda) &\simeq& \sum_{j}  \, x_{ij} \,  \tilde{C}^8_{ij} \, 
\frac{v_T^4}{\Lambda^4} + \sum_{j} \frac{(g^{ij}_{SM})^2}{16 \, \pi^2} \, \tilde{C}^6_{ij} \, y_{ij} \, 
\ln \left[\frac{\Lambda^2}{v_T^2}\right] \, \frac{v_T^2}{\Lambda^2} \spp
\eea
\end{linenomath}
Non log dependence in the second term is also present, but is suppressed for a simplifying 
approximation. Here $x_{ij},y_{ij}$ label the observable dependence and are $\mcO(1)$. 
One can further define 
\begin{linenomath}
\bea
x'_i \, \sqrt{N^{i}_8}  = \sum_{j} \sqrt{x^2_{ij} \, (\tilde{C}^8_{ij})^2},  \quad  
\quad y'_i \, \sqrt{N^i_6}  = \sum_{j} \sqrt{y^2_{ij} \,  (\tilde{C}^6_{ij})^2} \spc
\eea 
\end{linenomath}
as the product of $\mathcal{O}(1)$ numbers that characterize the multiplicity of the operators 
that contribute to a process ($N_{6,8}$) and the typical numerical dependence $x'_i,y'_i$.
The square root is because errors are assumed to add in quadrature. 
As an alternative, a Bayesian uniform prior for the $C_i$ could be used.

Although the number of 
operators is large, the relevant number of operators that contribute in a process is far less 
then the full operator set;
in known examples $N_{6,8} \sim \mcO(10)$. 
No complete operator basis of $\mcL_8$ has ever been encoded in a Monte-Carlo program 
and used to fit the data, and we do not recommend that 
fits should explicitly include all terms in $\mathcal{L}_8$ and vary corrections in general. Rough error estimates of this form should be sufficient for most purposes.

This error is multiplicative and the absolute 
error is obtained as $\Delta_{SMEFT}^i(\Lambda)$ times the SM prediction for an observable.
For cut off scales and Wilson coefficients in the range $1 \, {\rm TeV} \lesssim \Lambda/\sqrt{\tilde{C}_i} \lesssim 3 \, {\rm TeV}$ and order one 
numbers for $x_i,y_i,N_{6,8}$ the value of $\Delta_{SMEFT}^i(\Lambda)$
is in the range of few $\mcO(\%)$ to  
$\mcO(0.1\%)$ \cite{Berthier:2015gja,Berthier:2015oma,David:2015waa}. 
This is the reason we stress that once experimental errors descend to the $\mcO(10 \%)$ level SMEFT theory errors should
be considered to be conservative.  It is widely considered to be the case that the precision expected in LHC analyses can be 
expected to approach a few percent in well measured channels \cite{CMS:2013xfa,Flechl:2015foa}. 

The simplest approach to adopt is that a percentage error can be motivated for SMEFT theoretical uncertainties using these approximations and
then directly applied (and varied) when reporting a bound.

\subsection{NLO SMEFT loop corrections}

Including loop corrections in the SMEFT context is more crucial than for a pure SM calculation.
One loop corrections can introduce a dependence on Wilson coefficients that
do not contribute at tree level to a particular process and some of these Wilson coefficients are very poorly bounded.
This is different from the SM where all of the Lagrangian terms are extremely well known.
We will refer to the introduction of such dependence as ``non-factorizable'' corrections. Such corrections can significantly change
the interpretation of a mapping of experimental constraints at NLO in the SMEFT, as we illustrate below.
Loop corrections also introduce a perturbative rescaling of the dependence on an operator's 
Wilson coefficient. These corrections help define the variance discussed Sec. \ref{variancesection} for a LO analysis.

Improving the SMEFT to one loop requires 
a renormalization scheme be defined, a systematic
renormalization of the SMEFT be carried out on the new parameters in $\mathcal{L}_6$, and 
loop corrections be performed in a particular chosen gauge. 
We now discuss each of these steps in the NLO program in more detail.

\subsection{SMEFT: renormalization in practice}\label{inpractice}
In this Section we describe a general renormalization procedure in the SMEFT. 
The results 
presented have been developed in \Brefs{Passarino:2012cb,Ghezzi:2015vva}, based on the 
conventional formalism widely used in the SM \cite{'tHooft:1972ue,'tHooft:1972fi,'tHooft:1978xw,Passarino:1978jh}. 
To perform renormalization in an EFT it is appropriate to use a dimensionless regulator, see 
Refs.~\cite{Manohar:1996cq} for a review. We work with dimensional regularization and define
\begin{linenomath}
\bq
\DUV = \frac{2}{4 - d} - \emc - \ln \pi - \ln\frac{\muRs}{\mu^2} \spc
\eq
\end{linenomath}
where $d$ is space-time dimension, the loop measure is $\mu^{4 - d}\,d^nq$
and $\muR$ is the renormalization scale; $\emc$ is the Euler-Mascheroni constant. Counter-terms
for SM parameters and fields are defined by
\begin{linenomath}
\bq
\mrZ_i = 1 + \frac{g^2}{16\,\pi^2}\,\lpar d\mrZ^{(4)}_i + g_6\,d\mrZ^{(6)}_i \rpar\,\DUV \spp
\eq
\end{linenomath}
With field/parameter counter-terms we can make UV finite the self-energies and the 
corresponding Dyson resummed propagators. However, these counterterm subtractions are not 
enough to make UV finite the Green's functions with more than two legs 
(at $\mcO(g^{\mrN_6} g_6)$). A mixing matrix among Wilson coefficients is needed:
\begin{linenomath}
\bq
C_i = \sum_j\,\mrZ^{\sPW}_{ij}\,C^{\ren}_j \spc
\qquad 
\mrZ^{\sPW}_{ij} = \delta_{ij} + \frac{g^2}{16\,\pi^2}\,d\mrZ^{\sPW}_{ij}\,\DUV \spp
\eq
\end{linenomath}

For example, in this way we can renormalize the (on-shell) $\mrS\,$-matrix for
$\PH(P) \to \PA_{\mu}(p_1)\PA_{\nu}(p_2)$ and
$\PH(P) \to \PA_{\mu}(p_1)\PZ_{\nu}(p_2)$ which have only one (transverse) Lorentz structure.
By on-shell $\mrS\,$-matrix for an arbitrary process (involving unstable particles) we mean 
the corresponding (amputated) Green's function supplied with LSZ factors and sources, computed 
at the (complex) poles of the external lines~\cite{Grassi:2000dz,Kniehl:2001ch,Goria:2011wa}. For 
processes that involve stable particles this can be straightforwardly transformed into a 
physical observable.

The connection of the $\PH\PV\PV, \PV = \PZ,\PW$ (on-shell) $\mrS\,$-matrix with the off shell 
vertex $\PH \to \PVV$ and the full process $\Pp\Pp \to 4\,\psi$ is more complicated and 
is discussed in some detail in Sect.~3 of \Bref{David:2015waa}.
The ``on-shell'' $\mrS\,$-matrix for $\PH\PV\PV$, being built with the the residue of the 
$\PH{-}\PV{-}\PV$ poles in $\Pp\Pp \to 4\, \psi$ is gauge invariant by construction (it can be 
proved by using Nielsen identities~\cite{Grassi:2001bz}) and represents one of the building 
blocks for the full process: in other words, it is a 
pseudo-observable~\cite{Passarino:2010qk,Gonzalez-Alonso:2014eva,David:2015waa}.
Technically speaking the ``on-shell'' limit for external legs should be understood 
``to the complex poles'' (for a modification of the LSZ reduction formulas for unstable 
particles, see \Bref{Weldon:1975gu}) but, as well known, at one loop we can use 
on-shell masses (for unstable particles) without breaking the gauge parameter independence 
of the result. Residues of complex poles are what matters, as far as renormalization is concerned.

The $\PH(P) \to \PZ_{\mu}(p_1)\PZ_{\nu}(p_2)$ (on-shell) matrix contains a part of the amplitude
proportional to $g^{\mu\nu}$ (referred to as $\mcD_{\sPHZZ}$ below) and a part of the amplitude
proportional to $p^{\mu}_2\,p^{\nu}_1$ (referred to as $\mcP_{\sPHZZ}$ below). 
Both of these terms get renormalized through a mixing.

Consider now the $\PH(P) \to \PWm_{\mu}(p_1)\PWp_{\nu}(p_2)$ (on-shell) matrix:
it has the same Lorentz decomposition of $\PH \to \PZ\PZ$ and it is UV finite in the
$\mrdim = 4$ part. The $\mcD_{\sPHWW}$ part at $\mrdim = 6$ is renormalized through a mixing; 
however, there are no Wilson coefficients in $\mcP_{\sPHWW}$ that are not also present in  
$\mcP_{\sPHZZ}$, so that the UV finiteness of this term is related by gauge symmetry to the 
renormalization of $\mcP_{\sPHZZ}$. This is the first part of the arguments
used in \Brefs{Passarino:2012cb,Ghezzi:2015vva} in proving closure of NLO SMEFT 
under renormalization.

The (on-shell) decays $\PH(P) \to \PQb(p_1)\PAQb(p_2)$ and 
$\PZ(P) \to \bar{\psi}(p_1)\psi(p_2)$ are more involved to improve to NLO in the SMEFT. 
The SM contribution to these amplitudes are rendered finite by the SM counter-terms,
however renormalizing the contributions due to $\mcL_6$ requires an extensive 
treatment of this operator mixing. See Ref. \cite{Gauld:2015lmb} for recent results on these decays.


Some structure present in the SM is not preserved when extending an analysis into the SMEFT.
Manifestly, processes that first appear at one loop in the SM can occur at tree level in 
the SMEFT, due to the presence of local contact operators.
However, some symmetries of the SM are preserved. For example, consider the universality of the 
electric charge. 
In pure QED there is a Ward identity \cite{Ward:1950xp} telling us that $e$ can be renormalized in terms of 
vacuum polarization (which is a way to understand the universality of the coupling), and Ward-Slavnov-Taylor (WST) identities \cite{Ward:1950xp,Slavnov:1972fg,Taylor:1971ff} allow us to generalize the argument 
to the full spontaneously broken SM symmetry group.
The previous statement means that the contribution from vertices (at zero momentum transfer) in 
the full SM exactly cancel those from (fermion) wave function renormalization factors. Therefore, 
by directly computing the vertex $A \, \bar{\psi} \, \psi$ (at $q^2 = 0$) and the $\PZ_\psi$ wave function 
factor in the SMEFT, one can directly prove (or check) that the WST identity is extended to the SMEFT at $\mathcal{L}_6$.
This is expected as the corresponding identities are the consequence of symmetries. However, 
this is technically non-trivial even after the previous steps in the renormalization program 
discussed above. Once (non-trivial) finiteness of this vertex is established, the finiteness 
of $\Pep\Pem \to \bar{\psi} \, \psi$ (including the four-point functions in the non resonant part) follows. 
This is the second part in proving closure of the NLO SMEFT under renormalization, using the arguments 
of \Brefs{Passarino:2012cb,Ghezzi:2015vva}.

At NLO one first has to render all SM and SMEFT 
parameters finite. Considering the arguments above, and the complete renormalization results of 
all the operators in $\mcL_6$ reported in 
Refs.~\cite{Alonso:2013hga,Grojean:2013kd,Jenkins:2013zja,Jenkins:2013wua}
in the Warsaw basis, this step in the NLO program has been accomplished. This result has not been established in any other basis to date.
The defining conditions of some alternate LO constructions seem to make a renormalization program impossible to carry out.

\subsection{Input parameter choices}
The detailed fixing of poles and residues that make up precise renormalization conditions require a lengthy discussion.
For detailed reviews in the case of the SM, see Refs.~\cite{Denner:1994xt,Denner:1991kt}. Below we summarize the
results of the finite renormalization in the relationship to the input observables.

\subsubsection{Using a `$G_F$-scheme' with $G_{F}$, $M_W$, $M_Z$ }
In the `$G_{F}\,$-scheme', one uses $\{G_{F}\,,\,\mw\,,\,\mz\}$ to fix terms in the Lagrangian. In this case, we write the following 
equation for the $g$ finite renormalization
\begin{linenomath}
\bq
g_{\ren} = g_{\myexp} + \frac{g^2_{\myexp}}{16\,\pi^2}\,\lpar
d\mcZ^{(4)}_g + g_6\,d\mcZ^{(6)}_g \rpar \spc
\label{gexp}
\eq
\end{linenomath}
where $g_{\myexp}$ will be expressed in terms of the Fermi coupling constant $G_{F}$.
Furthermore, $\ctw = \mw/\mz$. The $\mu\,$-lifetime can be written in the form
\begin{linenomath}
\bq
\frac{1}{\tau_{\mu}} = \frac{M^5_{\PGm}}{192\,\pi^3}\,\frac{g^4}{32\,M^4}\,
\lpar 1 + \delta_{\mu} \rpar \spp
\eq
\end{linenomath}
The radiative corrections are $\delta_{\mu} = \delta^{\PW}_{\mu} + \delta_{\mrG}$
where $\delta_{\mrG}$ is the sum of vertices, boxes etc and $\delta^{\PW}_{\mu}$ is due
to the $\PW$ self-energy. The renormalization equation becomes
\begin{linenomath}
\bq
\frac{G_{F}}{\srt} = \frac{g^2}{8\,M^2}\,\left\{
1 + \frac{g^2}{16\,\pi^2}\,\Bigl[ \delta_{\mrG} + \frac{1}{M^2}\,\Sigma_{\PWW}(0)
\Bigr]
\right\} \spc
\label{GFscheme}
\eq
\end{linenomath}
where we expand the solution for $g$
\begin{linenomath}
\bq
g^2_{\ren} = 4\,\srt\,G_{F}\,M^2_{\PW\,;\,\ssOS}\,\left\{
 1 + \frac{G_{F} M^2_{\PW\,;\,\ssOS}}{2\,\sqrt{2}\,\pi^2}\,
\Bigl[ \delta_{\mrG} + \frac{1}{M^2}\,\Sigma_{\PWW\,;\,\fin}(0) \Bigr]
\right\} \spp
\label{GFsol}
\eq
\end{linenomath}
Note that the non universal part of the corrections is given by
\begin{linenomath}
\bq
\delta_{\mrG} = \delta^{(4)}_{\mrG} + g_6\,\delta^{(6)}_{\mrG} 
\quad
\delta^{(4)}_{\mrG} = 6 + \frac{7 - 4\,\stws}{2\,\stws}\,\ln\ctws \spc
\eq
\end{linenomath}
but the contribution of $\mathcal{L}_6$ to muon decay at NLO is not available yet and
has not be included in the calculation. It is worth noting that \eqn{GFscheme} defines the finite 
renormalization in the $\{G_{F}\,,\,\mw\,,\,\mz\}$ IPS.
\subsubsection{The `$\alpha\;$scheme', using $\alpha, G_{\rm{F}},M_Z$ } 
This scheme uses the fine structure constant $\alpha$ and is based on using
$\{\alpha\,,\,G_{F}\,,\,\mz\}$ as the IPS. The new finite-renormalization equation is
\begin{linenomath}
\bq
g^2\,\stws = 4\,\pi\alpha\,\Bigl[ 1 - \frac{\alpha}{4\,\pi}\,\frac{\Pi_{\PAA}(0)}{\stws} \Bigr]
\spc
\eq
\end{linenomath}
where $\alpha = \alpha_{\myQED}(0)$ and $\Pi_{\PAA}$ defines the vacuum polarization. Therefore, 
in this scheme, the finite counter-terms are
\begin{linenomath}
\bq
g^2_{\ren} = g^2_{\sPA}\,\Bigl[ 1 + \frac{\alpha}{4\,\pi}\,d\mcZ_{g} \Bigr] \spc
\quad
\ctwr = \cth\,\Bigl[ 1 + \frac{\alpha}{4\,\pi}\,d\mcZ_{\ctw} \Bigr],  
\quad
M_{\ren} = \mzOS\,\cths\,\Bigl[ 1 + \frac{\alpha}{8\,\pi}\,d\mcZ_{\mw} \Bigr] \spc
\label{eqIPSb}
\eq
\end{linenomath}
where the parameters $\cth$ and $g_{\sPA}$ are defined by
\begin{linenomath}
\bq
g^2_{\sPA} = \frac{4\,\pi\,\alpha}{\sths}
\qquad
\sths = \frac{1}{2}\,\Bigl[ 1 - \sqrt{1 - 4\,\frac{\pi\,\alpha}{\srt\,G_{F}\,\mzsOS}}\Bigr] \spp
\label{defscipsb}
\eq
\end{linenomath}
The reason for introducing this scheme is that the $\mrS, \mrT$ and $\mrU$ parameters 
\Bref{Peskin:1990zt} have been originally given in the $\{\alpha\,,\,G_{F}\,,\,\mz\}$ 
scheme, and these input parameters are very well measured in the SM. 
When calculating processes involving photons final states, this scheme can be transparent to adopt.
For other processes, the $\{G_{F}\,,\,\mw\,,\,\mz\}$ scheme can be more appropriate, and 
is in wider use in the SM in higher order calculations. In the $\alpha\,$-scheme, after 
requiring that $\mzsOS$ is a zero of the real part of the inverse $\PZ$ propagator, we are 
left with one finite counterterm, $d\mcZ_{g}$. The latter is fixed by using $G_{F}$ and 
requiring that
\begin{linenomath}
\bq
\frac{1}{\srt}\,G_{F} = \frac{g^2}{8\,M^2}\,\Bigl\{ 1 + \frac{g^2}{16\,\pi^2}\,\Bigl[
                       \delta_{\mrG} + \frac{1}{M^2}\,\Delta_{\PWW}(0)
                     - \lpar d\mrZ_{\PW} + d\mrZ_{\mw} \rpar\,\DUV \Bigr] \Bigr\} \spc
\eq
\end{linenomath}
where we use the following relations for UV and finite renormalization,
\begin{linenomath}
\bq
g = g_{\ren}\,\lpar 1 + \frac{g^2_{\ren}}{16\,\pi^2}\,d\mrZ_{g}\,\DUV \rpar 
\qquad
g_{\ren} = g_{\sPA}\,\lpar 1 + \frac{\alpha}{8\,\pi}\,d\mcZ_{g} \rpar \spp
\eq
\end{linenomath}
Note that SM EW calculations available in literature generally use $G_{F}$ for the pure 
weak part or evolve $\alpha(0) \to \alpha(M)$ and use $\alpha(M)$ as the expansion parameter 
at the scale $M$. For a comprehensive discussion see Sect.~5.3 of \Bref{Binoth:2010xt}.

\subsection{Background field gauge}

Any well defined gauge can be used in a calculation, see Ref.~\cite{Leibbrandt:1987qv} for an 
excellent review on gauge fixing. 
There can be some advantage to organising a calculation in a manner that enforces relationships 
between counter terms due to gauge invariance. A technique that accomplishes 
this is known as the Background Field (BF) method~\cite{DeWitt:1967ub,Abbott:1981ke}. The idea is that 
fields are split into classical and quantum components and a gauge fixing term is added that 
maintains the gauge invariance of the classical background fields, while breaking the gauge
invariance of the quantum fields. 
Due to the resulting Ward identities, one finds the relations among the SM counter-terms~\cite{Denner:1994xt}.
The gauge fixing in the BF method can be imposed as in Ref.~\cite{Denner:1994xt,Einhorn:1988tc}.
Use of the background field method can make extending the WST
relations between counter-terms manifest and transparent, even when including the effects 
of $\mcL_6$. It is worth noting that the WST identities have been explicitly verified in the straightforward LO approach detailed in this note.
Proving such identities in any LO approach verifies the gauge-independence of the results.

Extending any gauge fixing procedure to the case of the SMEFT is subtle, due to the order 
by order redefinition of the fields that are gauged due to terms in $\mcL_{SMEFT}$. 
Optimally resolving the technical complications that result is a challenge.
These subtleties are some of the reasons it is difficult to directly 
modify computer programs that have been developed for automatic NLO calculations in the SM, to the case of the SMEFT.
The development of NLO SMEFT Monte-Carlo tools is still very much a work in progress. 

\section{Known results in the SMEFT to NLO}
Despite all of the challenges to advancing SMEFT results to NLO, progress in this area is rapid 
and steady. In this section we briefly sumarize some of these theoretical developments.

\subsubsection{Renormalization results}
The complete renormalization of the Warsaw basis was reported in 
Refs.~\cite{Alonso:2013hga,Grojean:2013kd,Jenkins:2013zja,Jenkins:2013wua}.
In the approach outlined in Section \ref{inpractice}, results for the Warsaw basis
operator renormalization were reported in \Brefs{Passarino:2012cb,Ghezzi:2015vva}. 
Use of SMEFT renormalization results (including a subset of NLO finite terms) to leverage 
EWPD to bound operators not contributing at tree level was reported in Ref.~\cite{deBlas:2015aea}.
Partial results for renormalizing some alternate operator sets in a so called ``SILH basis''
were given in Refs.~\cite{Elias-Miro:2013eta,Elias-Miro:2013mua}. A recent study of RGE effects on the
oblique parameters, in a subset of UV models, was reported in Ref.\cite{Wells:2015cre}.

\subsubsection{Advances in one loop matching techniques}
Recently, the covariant derivative expansion discussed in Refs.~\cite{Cheyette:1985ue,Cheyette:1987qz,Gaillard:1985uh} has re-emerged
in Refs.\cite{Henning:2014wua,Drozd:2015kva,Drozd:2015rsp} as a powerful technique to perform matching calculations
to underlying UV theories at one loop. The basic idea at work is that, the contribution to the effective action that results when integrating out a heavy field $\mathcal{X}$ 
at one loop is schematically given by 
\begin{linenomath}
\bea
\Delta S \propto i \, {\rm Tr} \, \log \left[D^2+ m_\mathcal{X}^2 + U(x) \right]
\eea
\end{linenomath}
where $m_\mathcal{X}$ is the mass of the $\mathcal{X}$ field integrated out, $D^2 = D_\mu \, D^\mu$, $D_\mu$ is the covariant derivative, 
and $U(x)$ depends on the SM field content. The covariant derivative expansion allows this functional trace to be directly
evaluated, while keeping gauge covariance manifest. This simplifies and systematizes one loop matching calculations in the SMEFT,
in many simple UV physics cases.\footnote{It is worth noting, that some questions remain about the effect of mixing between the heavy and light field content in this approach \cite{Brehmer:2015rna}. These questions were recently clarified in \cite{delAguila:2016zcb}.}

\subsubsection{Full Lagrangian expansion results to NLO $(\mathcal{L}_8$)}

Refs.\cite{Lehman:2014jma,Henning:2015alf,Lehman:2015coa,Lehman:2015via,Henning:2015daa}
have developed the theoretical technology (essentially advanced use of Hilbert series techniques) to characterize the number of
independent operators present at each order in the SMEFT expansion. This has lead to the complete characterization of the operator
sets in $\mathcal{L}_7$ and $\mathcal{L}_8$.

\subsubsection{Perturbative NLO results in the SMEFT}

Full results to NLO in the SMEFT have started to appear in the literature. The first pioneering calculations of this
form were for the process $\mu \rightarrow e \, \gamma$ in Ref.\cite{Pruna:2014asa} and for the process
$\Gamma(\PH \rightarrow \PGg \, \PGg)$ in Refs.~\cite{Hartmann:2015oia,Hartmann:2015aia,Ghezzi:2015vva}.
In \cite{Hartmann:2015aia} the full NLO perturbative SMEFT result for 
this decay with no assumption
in the underlying UV scenario was reported. Ref.~\cite{Ghezzi:2015vva} also reported NLO results for 
$\Gamma(\PH \rightarrow \PZ \, \PGg)$, $\PH \rightarrow \PZ \, \PZ^\star$,
$\PH \rightarrow \PW \, \PW^\star$ under the assumption of a PTG scenario and presented
results to NLO for the $\PW$ mass and other EWPD parameters. Recently Ref.~\cite{Gauld:2015lmb} 
also reported NLO perturbative results for $\PH\rightarrow \PAQb\PQb$ and 
$\PH \rightarrow \PGtm\PGtp$  in the general SMEFT, including finite terms, in the 
large $m_{\PQt}$ limit.  NLO QCD results for a set of higher dimensional operators contributing to the Higgs pair production process
were given in Ref.\cite{Grober:2015cwa}, for the Higgs characterization model in Ref.\cite{Artoisenet:2013puc} and for associated Higgs production in
Ref.\cite{Mimasu:2015nqa}.

\subsection{A study of constraints} 
As a particular example, we
discuss the impact of NLO corrections on inferred LO bounds, in the case of 
$\Gamma(\PH \rightarrow \PGg \, \PGg)$, using the results of 
Refs.~\cite{Hartmann:2015oia,Hartmann:2015aia}.
We consider the general SMEFT case, consider unknown $\tilde{C}_i \sim 1$ and vary the unknown parameters 
over $0.8 \leq \Lambda \leq 3$ in ${\rm TeV}$ units. 
Note that $\bar{v}_T^2/(0.8 \, {\rm TeV})^2 \sim 0.1$. Taking $\kappa_{\gamma}$ from 
Ref.~\cite{ATLAS:2015bea} to be $0.93^{+0.36}_{-0.17}$, 
and neglecting light fermion ($m_{\Pf} < m_{\PH}$) effects for simplicity, one finds
the $1 \, \sigma$ range 
\begin{linenomath}
\bea\label{masternumerics}
-0.02 \, \leq \left(\tilde{C}^{1,NP}_{\PGg\,\PGg} + \frac{\tilde{C}^{NP}_i\,f_i}{16\,\pi^2}\right) 
\, \frac{\bar{v}^2_T}{\Lambda^2} \leq 0.02 \spp
\eea
\end{linenomath}
Here, the tilde superscript denotes that the scale $1/\Lambda^2$ has been factored out of a 
Wilson coefficient. The $f_i$ terms correspond to the ``nonfactorizable'' terms, and $\tilde{C}^{1,NP}_{\PGg\,\PGg}$
corresponds to the one loop improvement of the Wilson coefficient that gives this decay at tree level -- $\tilde{C}^{0,NP}_{\PGg \, \PGg}$.
The difference in the mapping of this constraint to the coefficient of $\tilde{C}^{0,NP}_{\PGg \, \PGg}$ 
at tree level, and at one loop, can now be characterized.

To determine this correction we determine the percentage change on the inferred
value of the bounds of $\tilde{C}^{0,NP}_{\PGg \, \PGg}$, while shifting the quoted upper and 
lower experimental bounds by the NLO SMEFT perturbative correction. The 
envelope of the two percentage variations on the bounds is quoted in the form $\left[ ,\right]$, 
for values of $\Lambda$ varying from $\left[0.8, \, 3\right] {\rm TeV}$.
For one specific choice of signs for $C_i$, we find the following characteristic results.
The net impact of one-loop corrections (added in quadrature) due to higher dimensional 
operators on the bound of the tree level Wilson coefficient  is
\begin{linenomath}
\bea
\Delta_{\text{quad}} \,  \tilde{C}^{0,NP}_{\PGg \, \PGg} \sim \left[29, 4 \right]  \% \spp
\eea
\end{linenomath}
Similarly, CMS reports $\kappa_{\gamma}= 0.98^{+0.17}_{-0.16}$ \cite{Khachatryan:2014jba}, 
which gives
\begin{linenomath}
\bea
\Delta_{\text{quad}} \,  \tilde{C}^{0,NP}_{\PGg \, \PGg} \sim \left[52, 7 \right]  \% \spp
\eea
\end{linenomath}
It is possible that these corrections could add up in a manner that is not in quadrature, as 
this depends on the unknown $\tilde{C}_i$ values. The impact of the one-loop corrections listed 
above is on {\it current} experimental bounds of $\Gamma(\PH \rightarrow \PGg \PGg)$, following from our conservative
treatment of unknown UV effects. 
As the experimental precision of the measurement of $\Gamma(\PH \rightarrow \PGg \PGg)$ 
increases, the impact of the neglected corrections directly scales up. Repeating the exercise 
above, with a chosen projected RunII value $\kappa_{\gamma} = 1 \pm 0.045$ which is consistent with projected 
future bounds (CMS - scenario II~\cite{Flechl:2015foa,CMS:2013xfa})
\begin{linenomath}
\bea
(\Delta_{\text{quad}} \, \tilde{C}^{0,NP}_{\PGg \, \PGg})^{\text{proj:RunII}} \sim 
\left[167, 21 \right]  \% \spp
\eea
\end{linenomath}
High luminosity LHC runs are further quoted to have a sensitivity between $2 \%$ and $5 \%$ in 
$\kappa_{\gamma}$ \cite{Dawson:2013bba}. Choosing a value 
$\kappa_{\gamma} = 1 \pm 0.03$ for this case, one finds 
\begin{linenomath}
\bea
(\Delta_{\text{quad}} \, \tilde{C}^{0,NP}_{\PGg \, \PGg})^{\text{proj:HILHC}} \sim 
\left[250, 31 \right]  \% \spp
\eea
\end{linenomath}
Neglected one loop corrections can have an important effect on the projection of an
experimental bound into the LO SMEFT formalism, when measurements become sufficiently precise and
the cut off scale is not too high.

\subsection{A study of SM-deviations}
Here the reference process is the off-shell $\Pg\Pg \to \PH$ production. 
It is important to go off-shell because the correct use of the SMEFT proves that scaling couplings 
on a resonance pole is not the same thing as scaling them off of a resonance pole, which has important
consequences in bounding the Higgs intrinsic width, see
\Brefs{Englert:2015bwa,Englert:2015zra,Ghezzi:2014qpa}.

In the $\upkappa\,$ approach, which was developed out of Refs.\cite{Azatov:2012bz,Espinosa:2012ir,Carmi:2012yp}, and formalized in 
Ref.~\cite{LHCHiggsCrossSectionWorkingGroup:2012nn}, one writes the amplitude as
\begin{linenomath}
\bq
\mrA^{\sPgg} = \sum_{\PQq=\PQt,\PQb}\,\upkappa^{\sPgg}_{\PQq}\,
\mcA^{\sPgg}_{\PQq} + \upkappa^{\sPgg}_c \spc
\label{kappafact}
\eq
\end{linenomath}
$\mcA^{\sPgg}_{\PQt}$ being the SM $\PQt\,$-loop \etc The contact term (which is the 
LO SMEFT) is given by $\upkappa^{\sPgg}_c$. Furthermore
$\upkappa^{\sPgg}_{\PQq} = 1 + \Delta\,\upkappa^{\sPgg}_{\PQq}$
Next we compute the following ratio
\begin{linenomath}
\bq
\mrR = 
\sigma\lpar \upkappa^{\sPgg}_{\PQq}\,,\,\upkappa^{\sPgg}_c \rpar/\sigma_{\mySM} - 1 \quad  [\%]
\spp
\eq
\end{linenomath}

\begin{figure}[t]
   \centering
   \includegraphics[width=0.7\textwidth, trim = 30 250 50 80, clip=true]{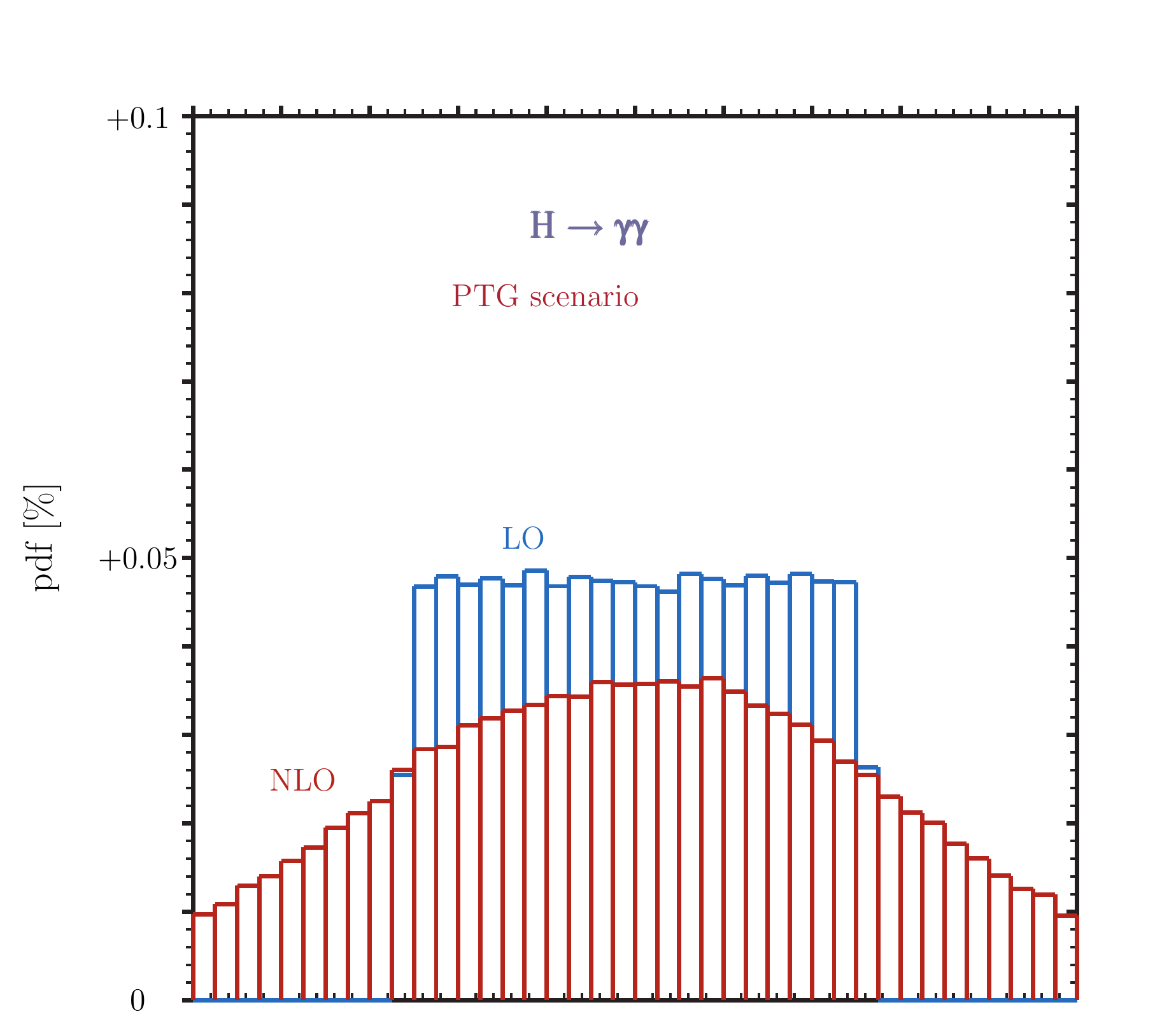}
\caption[]{Probability distribution function for the decay $\PH \to \PGg\PGg$ with a comparison 
between the LO and the NLO predictions. Here $\Lambda = 3\UTeV$ and $n = 1$. X axis as in previous figures.}
\label{HAApdf}
\end{figure}

In LO SMEFT $\upkappa_c$ is non-zero and $\upkappa_{\PQq} = 1$. One measures a deviation and gets 
a value for $\upkappa_c$.  However, at NLO $\Delta \upkappa_{\PQq}$ is non zero and one gets a 
degeneracy: the interpretation in terms of $\upkappa^{\myLO}_c$ or in terms of 
$\{\upkappa^{\myNLO}_c, \Delta \upkappa^{\myNLO}_{\PQq}\}$ could be rather different
(we show an example in Fig.~\ref{HAApdf}). Going interpretational we consider
\begin{linenomath}
\bq
\mrA^{\sPgg}_{\EFT} = 
\frac{g\,g_3}{\pi^2}\,\sum_{\PQq=\PQt,\PQb}\,\upkappa^{\sPgg}_{\PQq}\,\mcA^{\sPgg}_{\PQq}
+   2\, g_3 \,g_{_6}\,\frac{s}{\mws}\, \tilde{C}_{\PH\,\Pg}  + 
    \frac{g\,g_3\,g_{_6}}{\pi^2}\,\sum_{\PQq=\PQt,\PQb}\,\mcA^{\nfct\,;\,\sPgg}_{\PQq}\,
      \tilde{C}_{\PQq \Pg} \spc
\label{SMEFTfact}
\eq
\end{linenomath}
where $g_3$ is the $SU(3)$ coupling constant. Using \eqn{SMEFTfact} we adopt the Warsaw basis 
and eventually work in the (PTG) scenario~\cite{Arzt:1994gp,Einhorn:2013kja}. The following options are available: 
LO SMEFT: $\upkappa_{\PQq} = 1$ and $\tilde{C}_{\PH\,\Pg} $ is scaled by $1/16\,\pi^2$ 
being ``loop-generated'' (LG); 
NLO PTG-SMEFT: $\upkappa_{\PQq} \not= 1$ but only PTG operators 
inserted in loops (non-factorizable terms absent), $\tilde{C}_{\PH\,\Pg}$  scaled as above;
NLO full-SMEFT: $\upkappa_{\PQq} \not= 1$ LG/PTG operators 
inserted in loops (non-factorizable terms present), LG coefficients scaled as above.
Again we note the PTG  classification scheme is not valid for all
possible UV.

It is worth noting the difference between \eqn{kappafact} and \eqn{SMEFTfact}, showing that the
original $\upkappa\,$-framework can be made consistent at the price of adding 
``non-factorizable'' sub-amplitudes. At NLO, $\Delta \upkappa = g_{_6}\,\rho$ and
\begin{linenomath}
\bqa
g^{-1}_{_6} = \sqrt{2}\,G_{F}\,\Lambda^2
&\qquad&
4\,\pi\,\alphas = g_3 \spc \\
\rho^{\sPgg}_{\PQt} = \tilde{C}_{\PH\,\PW} + \tilde{C}_{\PQt\PH}  + 2\, \tilde{C}_{\PH\,\Box} - 
\frac{1}{2}\, \tilde{C}_{\PH\,\PD}
&\qquad&
\rho^{\sPgg}_{\PQb} = \tilde{C}_{\PH\,\PW} - \tilde{C}_{\PQb\PH}+ 2\, \tilde{C}_{\PH\,\Box} - 
\frac{1}{2}\, \tilde{C}_{\PH\,\PD} \spp \nn
\eqa
\end{linenomath}
Relaxing the PTG assumption introduces non-factorizable sub-amplitudes proportional to
$\tilde{C}_{\PQt\PH} , \tilde{C}_{\PQb\PH} $ with a mixing among $\tilde{C}_{\PH\,\Pg}, \tilde{C}_{\PQt \Pg}, 
\tilde{C}_{\PQb \Pg}$. Meanwhile, renormalization 
has made one-loop SMEFT finite, \eg in the $G_{F}\,$-scheme, with a residual 
$\muR\,$-dependence.

\begin{figure}[t]
\includegraphics[scale=0.37]{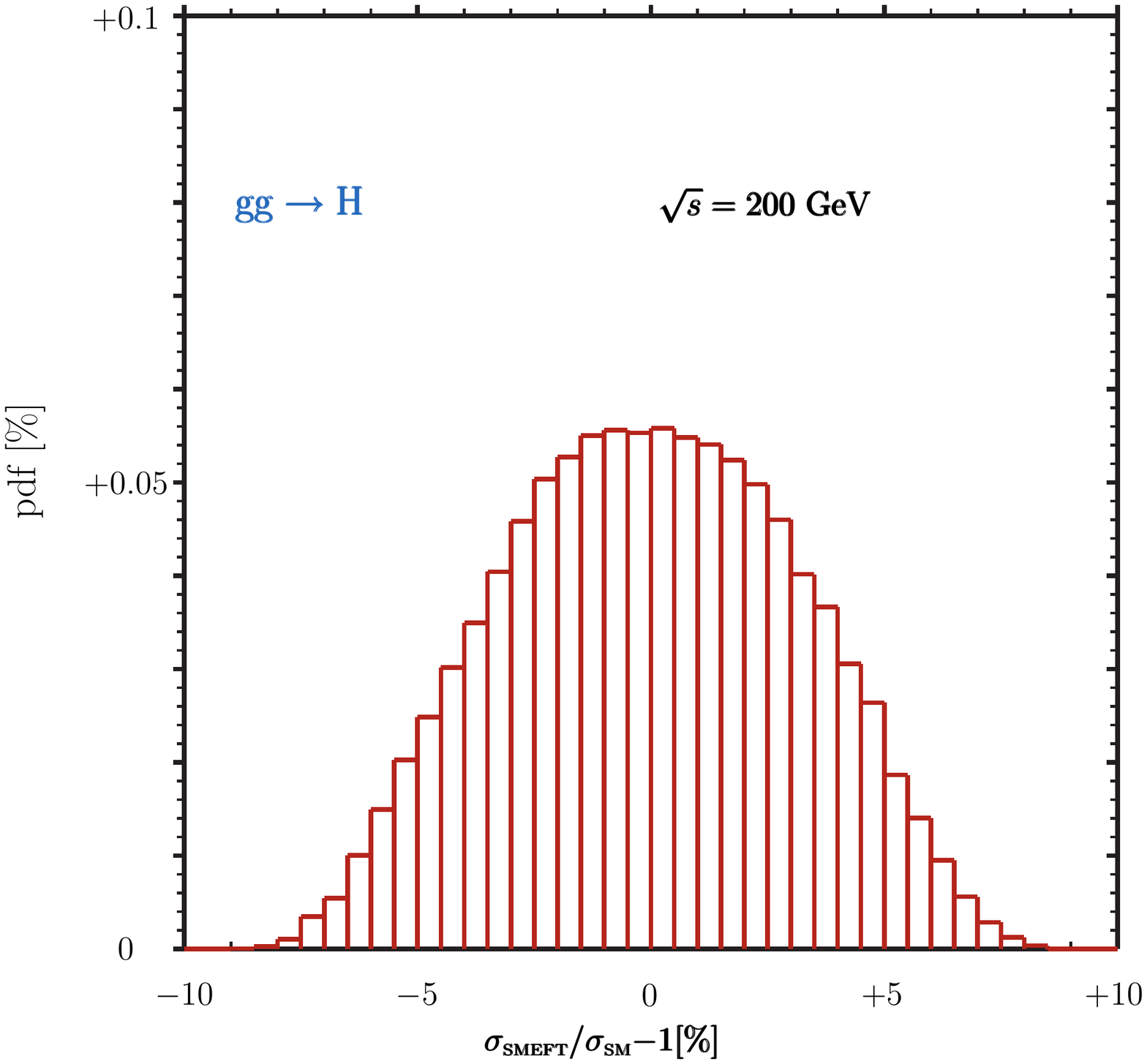}
\includegraphics[scale=0.37]{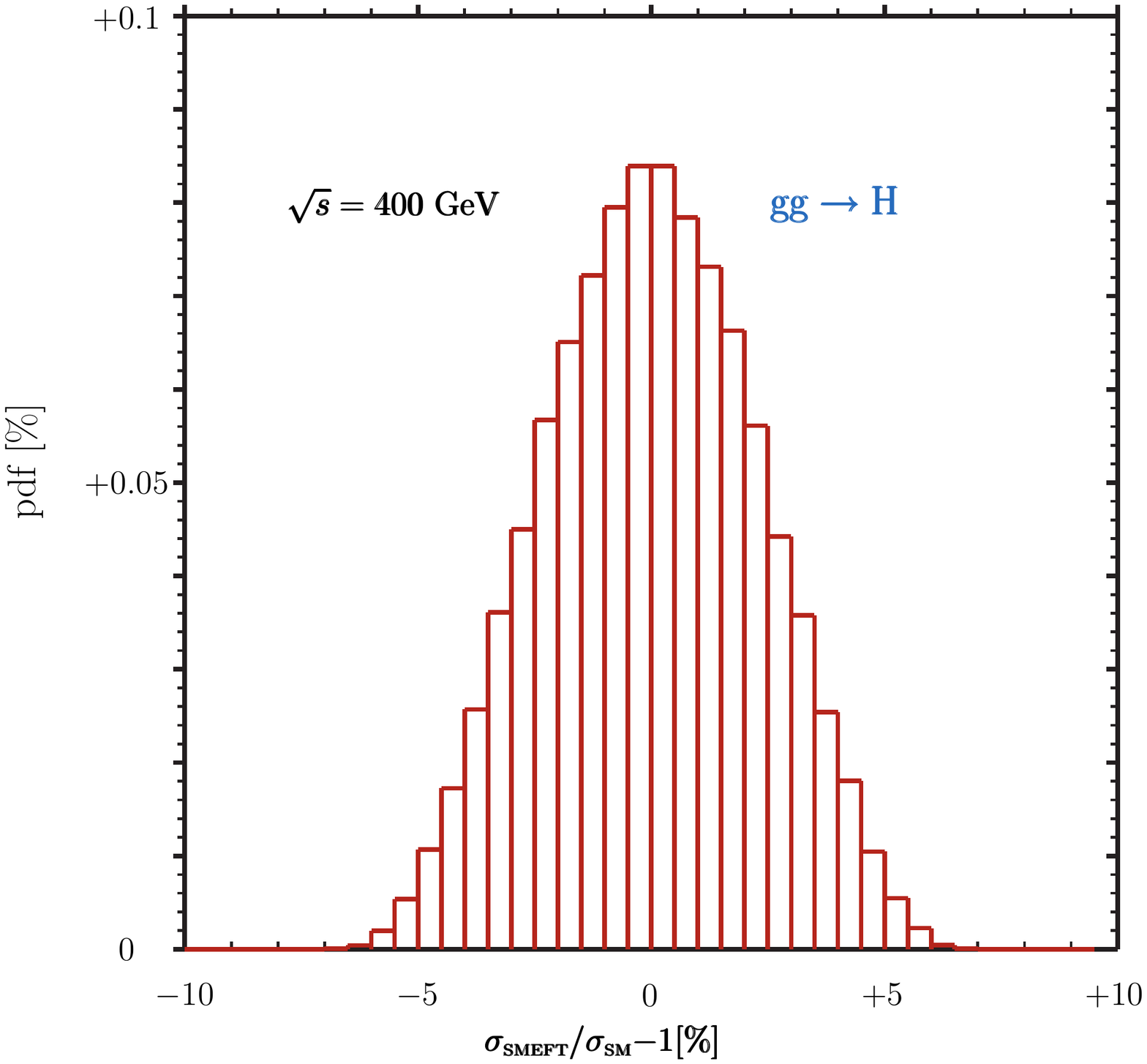}



\caption{Probability distribution function for the off-shell process $\Pg\Pg \to \PH$. Support is $C_i\;\in\;[-1\,,\,+1]$
with a uniform prior, and we have set $\Lambda = 3 \, {\rm TeV}$.}
\label{figurepdf2}
\end{figure}

\begin{figure}[t]
\includegraphics[scale=0.32,angle=90]{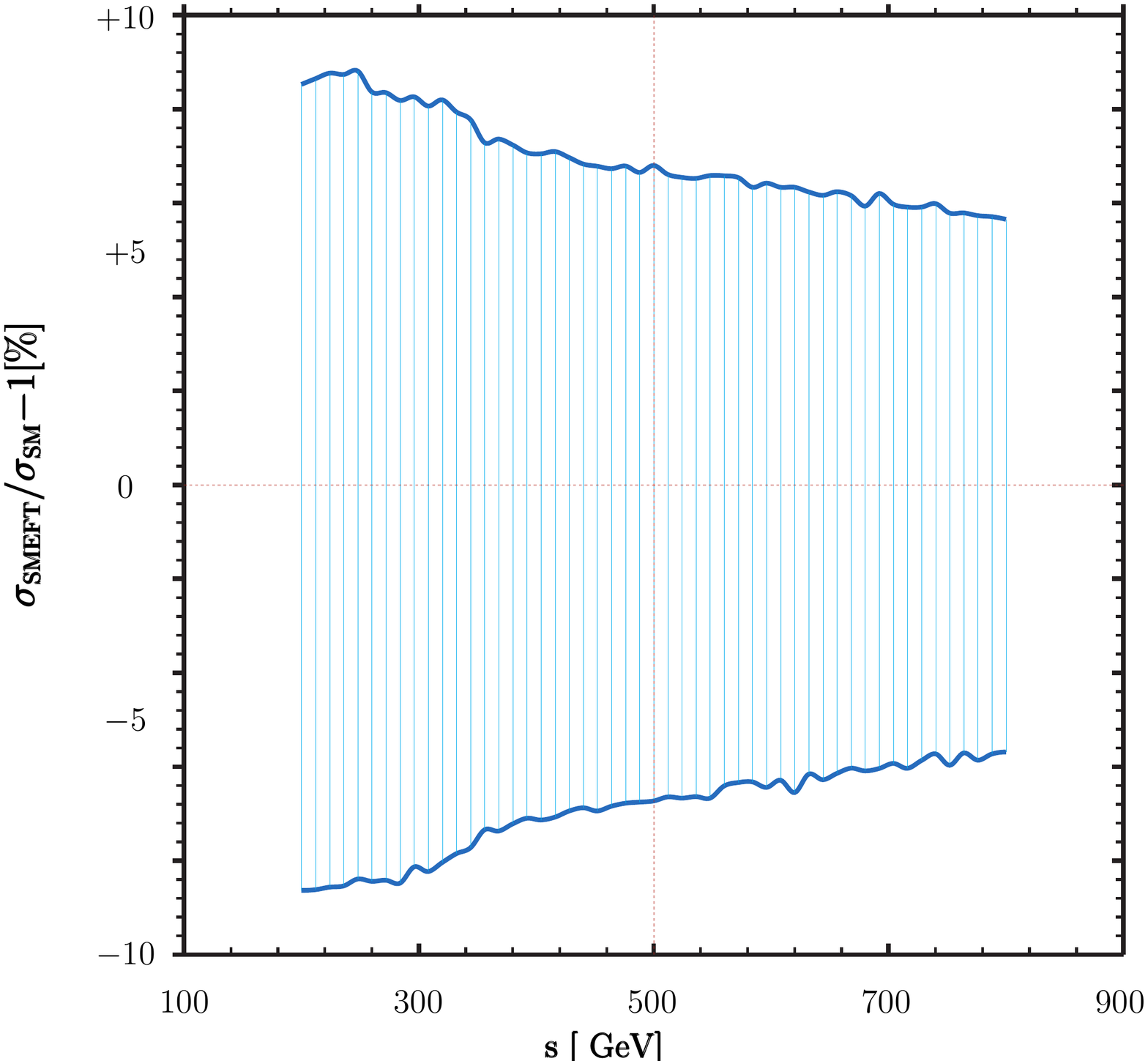}
\put(-15,-45){\includegraphics[scale=0.33]{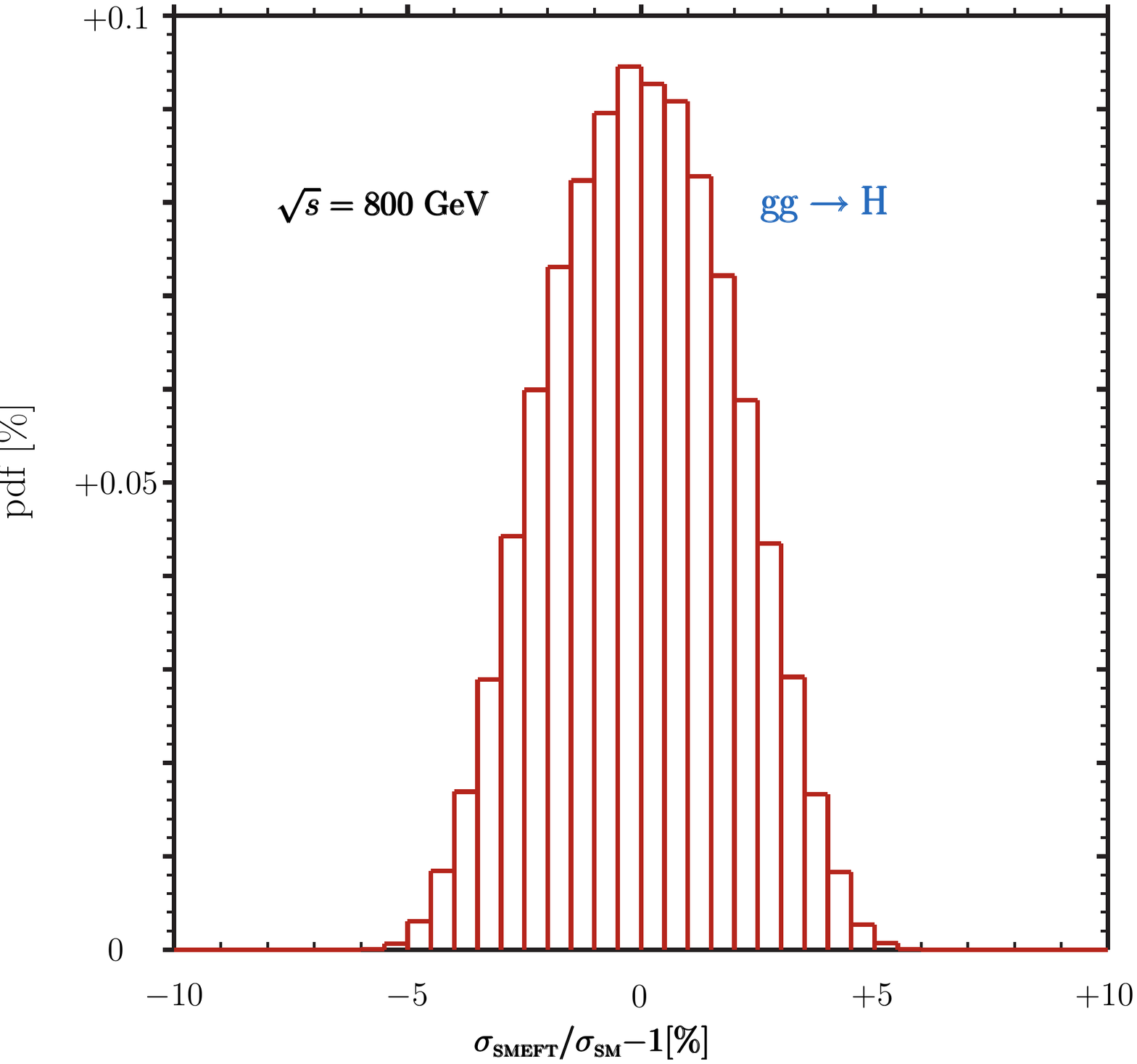}}
\caption{Probability distribution function for the off-shell process $\Pg\Pg \to \PH$. Support is $C_i\;\in\;[-1\,,\,+1]$
with a uniform prior, and we have set $\Lambda = 3 \, {\rm TeV}$.}
\label{figurepdf}
\end{figure}

We allow each Wilson coefficient to vary in some interval $\mrI_n = [-n\,,\,+n]$ and fix a 
value for $\Lambda$. Next we generate points from $\mrI_n$ for the Wilson coefficients 
with uniform probability and calculate $\mrR$. Finally, we calculate the $\mrR$ probability
distribution function (pdf), as shown in Figs.~\ref{figurepdf2},\ref{figurepdf}.

As another example, a comparison between the LO pdf and NLO pdf for $\PH \to \PGg\PGg$
using the approach of this section, and the results in \cite{Ghezzi:2015vva}, is shown in Fig.~\ref{HAApdf}.

\subsection{Comments on Pole observables vs tails of distributions}

When analyzing data near poles, scaling arguments that apply to the suppression of local contact (non resonant) four fermion operators in $\mcL_6$ 
also apply to NLO $\mcL_8$ corrections. This is fortunate as the very large number of parameters present in 
$\mcL_8$ and $\mcL_6$ are primarily present in four fermion operators. In the 
case of $\mcL_6$ $2205$ of the $2499$ parameters present are due to four fermion operators \cite{Alonso:2013hga}.  NLO power corrections in $\mcL_8$, higher order in $(v/\Lambda)^m$, are suppressed compared to $\mcL_6$ by the 
power counting parameter $v^2/\Lambda^2$,which
varies from $\sim 6\%$ to $\sim 0.6\%$ for $\Lambda/\sqrt{\tilde{C_i}} = 1,3 \, {\rm TeV}$ respectively.

The suppression of NLO terms in the Lagrangian expansion that scale as $p^2/\Lambda^2$ can be far less in the tails of distributions\footnote{See for example discussion in Ref.\cite{Ellis:2012xd,Isidori:2013cga,Biekoetter:2014jwa}.}. 
Tails of distributions can also have a very large number of  SMEFT parameters contributing due to non-resonant fermion pair (and higher multi-body) production background
processes. The SMEFT 
expansion breaks down when $p^2/\Lambda^2 \sim 1$, and Pseudo Observable/form factor \cite{David:2015waa,Passarino:2010qk,Gonzalez-Alonso:2014eva,Isidori:2013cla,Isidori:2013cga}
methods are required to characterize the data in this case.  In doing so, it is appropriate to bin the data in a manner that is transparent as to the momentum scale being probed.

It is also worth noting that unlike the case of pole data, NLO corrections to tails of distributions
are complicated in their analysis, as the $p^2/\Lambda^2$ terms are in general not gauge invariant alone, and need to always be combined
with the interference with non-resonant part of the SM, and SMEFT background processes. The requirement for joint analysis
including SMEFT corrections on the background that results, further complicates the analysis of non-pole data.

\section{Summary  and comments}
The takeaway points are as follows.
\begin{itemize}
\item{NLO results have already had an important impact on the SMEFT physics program. They
have shown that LEP constraints are not as strong as claimed in some literature, with bounds relaxed in the consistent
EFT interpretation from $\mathcal{O}(10^{-3}) \rightarrow \mathcal{O}(10^{-2})$. Due to this LEP constraints should not be interpreted to mean that 
effective SMEFT parameters in $\mcL_6$, or combinations of such parameters for vector boson couplings to fermions, should be 
set to zero in LHC analyses. Care should be used when fixing combinations of parameters
from EW constraints in LHC analyses. Arguments leading to claims of $\mathcal{O}(10^{-3})$ bounds are based on LO SMEFT analyses without any theoretical error assigned.}

\item{ It is important to preserve the original data, not 
just the interpretation results, as the estimate of the missing higher order terms can change over time,
modifying the lessons drawn from the data and projected into the SMEFT.} 

\item{Overall, the neglect of NLO (perturbative EW) corrections, considering the precision of LHC RunI measurements, 
is (retrospectively) justified in most channels. On the other hand, NLO QCD corrections are not neglectable, even in RunI.
However, considering projections for the precision to be reached in LHC RunII analyses, LO results 
for interpretations of the data in the SMEFT are challenged by consistency concerns on both of these fronts, and are not sufficient. This is particularly the case if the cut off scale
is in the few TeV range.}

\item{NLO results are starting to become available in the SMEFT. These results allow the 
consistent interpretation of the data combining measurements at different scales, and can 
robustly accommodate the precision projected to be achieved in RunII analyses, even for lower 
cut off scales.}

\item{In a sense, NLO results allow the kappa-framework~\cite{LHCHiggsCrossSectionWorkingGroup:2012nn} to be extended/replaced.
The idea is that interpretations can transition to the linear SMEFT, which is a systematically improvable  EFT formalism.
NLO results more consistently include kinematic deviations from the SM, 
and define higher order calculations in relation to a measured observable, in a well defined field theory.
A properly formulated SMEFT is not limited to LO and can include
QCD and EW corrections.} 

\item{The assignment of a theoretical error for LO SMEFT analyses is always important. This is essential if the cut off 
scale is assumed to be in the ``interesting range'' 
$1\, {\rm TeV} \lesssim \Lambda /\sqrt{\tilde{C_i}}  \lesssim 3\, {\rm TeV}$ and the experimental precision of analyses descends below the $10\%$ level.
The exact size of NLO corrections depends on the particular UV model, which is unknown, and also the particular channel
analysed.}

\item{Absorbing the effects of $\mathcal{L}_8$ corrections and/or absorbing logarithmic NLO perturbative corrections into an ``effective" parameter
to attempt to incorporate NLO corrections is very questionable. Such a redefinition cannot simultaneously be made in different measurements
generally measured at different scales. Correlating different measurements
is necessary if the SMEFT is to be used in a predictive fashion for constraints on LHC measurements.}

\item{We think that the experimental collaborations should restrict the bulk of their efforts to
defining and reporting clean measurements that can be interpreted in any well defined basis in the SMEFT.
The focus for data reporting should be on fiducial cross sections and/or pseudo-observables.
If a LO interpretation of the data in the SMEFT is reported there is no barrier to using the straightforward LO formalism of the Warsaw basis discussed in this note. This approach is convenient and well defined.}

\end{itemize}

We have supplied the outline and details of a LO implementation in Section \ref{rotation}. There is evidence enough to prove that the adoption of this
approach for LO fits is theoretically advantageous. We have sketched out how fits can be pursued at LO and NLO in a consistent fashion using this formalism.
The approach presented is well defined, is not intrinsically tied to a particular IPS, can be informed by theoretical errors determined at NLO and can be directly improved to NLO.
The gauge invariance of the approach presented has been checked at NLO by explicit confirmation of the WST identities.

We have stressed the standard usage of EFT terminology in this discussion, in particular the definition of an operator basis, to clarify discussion on these issues. 
EFT is traditionally a very successful paradigm to use to interpret the data because it is implemented as a well defined field theory.
Standard EFTs can be systematically improved from LO to NLO \xxxr{as they avoid ad-hoc and ill defined assumptions and Lagrangian manipulations.}
Very severe caution should be exercised when considering approaches that are presented as EFT that are not constructed in such a standard and well defined manner.

\small

\section*{Afterword}
In this document, we have highlighted in red some of the key text of this document that was erased in YR4  \cite{deFlorian:2016spz}. Other text was also significantly changed,
modified to have the opposite of its initial meaning, or also erased in the YR4 version of this document. The interested reader can compare the texts. The authors do not endorse these editorial actions.

\bibliographystyle{JHEP}
\bibliography{NLOSMEFT}

\end{document}

%% file: proleg.tex
\usepackage[dvipsnames]{xcolor}
\usepackage{cite}
\usepackage[scale={.75,.8}]{geometry} 
\usepackage{epic}
\usepackage{latexsym}
\usepackage{verbatim}
\usepackage{epsf}
\usepackage{epsfig}
\usepackage{xcolor}
\usepackage{ulem} 

\usepackage{cancel}

\usepackage{amssymb}
\usepackage{heppennames2}
\usepackage{lhchiggs}
\usepackage{cernunits}
\usepackage[left]{lineno}
\DeclareSymbolFont{forjmath}{OT1}{cmr}{m}{sl}
\DeclareMathSymbol{\Jmath}{\mathord}{forjmath}{'021}
\def\jmath{\Jmath}
\DeclareFontFamily{OT1}{cmr}{}
\DeclareFontFamily{OT1}{cmss}{}
\usepackage{pifont}
\usepackage{fancybox}
\usepackage{braket,slashed,bm}
\usepackage[abs]{overpic}
\usepackage{parskip}
\usepackage{enumerate}
\usepackage{url}
\usepackage[utf8]{inputenc}
\allowdisplaybreaks

\makeatletter
\def\section{\@startsection{section}{0}{\z@}{5.5ex plus .5ex minus
 1.5ex}{2.3ex plus .2ex}{\large\bf}}
\def\subsection{\@startsection{subsection}{1}{\z@}{3.5ex plus .5ex minus
 1.5ex}{1.3ex plus .2ex}{\normalsize\bf}}
\def\subsubsection{\@startsection{subsubsection}{2}{\z@}{-3.5ex plus
-1ex minus  -.2ex}{2.3ex plus .2ex}{\normalsize\sl}}
\newcommand{\xxxr}[1]{{\color{red}{#1}}}

\def\nn{\nonumber }

\def\bea{\begin{eqnarray}}
\def\eea{\end{eqnarray}}
\renewcommand{\@makecaption}[2]{%
   \vskip 10pt
   \setbox\@tempboxa\hbox{\small #1: #2}
   \ifdim \wd\@tempboxa >\hsize     
       \small #1: #2\par          
     \else                        
       \hbox to\hsize{\hfil\box\@tempboxa\hfil}
   \fi}
 \def\citenum#1{{\def\@cite##1##2{##1}\cite{#1}}}
\def\citea#1{\@cite{#1}{}}
\newcount\@tempcntc
\def\@citex[#1]#2{\if@filesw\immediate\write\@auxout{\string\citation{#2}}\fi
  \@tempcnta\z@\@tempcntb\m@ne\def\@citea{}\@cite{\@for\@citeb:=#2\do
    {\@ifundefined
       {b@\@citeb}{\@citeo\@tempcntb\m@ne\@citea\def\@citea{,}{\bf }\@warning
       {Citation `\@citeb' on page \thepage \space undefined}}%
    {\setbox\z@\hbox{\global\@tempcntc0\csname b@\@citeb\endcsname\relax}%
     \ifnum\@tempcntc=\z@ \@citeo\@tempcntb\m@ne
       \@citea\def\@citea{,}\hbox{\csname b@\@citeb\endcsname}%
     \else
      \advance\@tempcntb\@ne
      \ifnum\@tempcntb=\@tempcntc
      \else\advance\@tempcntb\m@ne\@citeo
      \@tempcnta\@tempcntc\@tempcntb\@tempcntc\fi\fi}}\@citeo}{#1}}
\def\@citeo{\ifnum\@tempcnta>\@tempcntb\else\@citea\def\@citea{,}%
  \ifnum\@tempcnta=\@tempcntb\the\@tempcnta\else
  {\advance\@tempcnta\@ne\ifnum\@tempcnta=\@tempcntb \else\def\@citea{--}\fi
    \advance\@tempcnta\m@ne\the\@tempcnta\@citea\the\@tempcntb}\fi\fi}
\makeatother
\newcommand{\eqn}[1]{Eq.(\ref{#1})}

\newcommand{\Bref}[1]{Ref.~\cite{#1}}
\newcommand{\Brefs}[1]{Refs.~\cite{#1}}

\newcommand{\lpar}{\left(}                            
\newcommand{\rpar}{\right)} 

\newcommand{\spc}{\,,}
\newcommand{\spp}{\,.}
\newcommand{\bq}{\begin{equation}}                   
\newcommand{\eq}{\end{equation}}
\newcommand{\bqa}{\arraycolsep 0.14em\begin{eqnarray}}
\newcommand{\eqa}{\end{eqnarray}}
\newcommand{\mrs}{{\mathrm{s}}}
\newcommand{\mrc}{{\mathrm{c}}}
\newcommand{\mrA}{{\mathrm{A}}}

\newcommand{\mrD}{{\mathrm{D}}}

\newcommand{\mrG}{{\mathrm{G}}}
\newcommand{\mrI}{{\mathrm{I}}}

\newcommand{\mrN}{{\mathrm{N}}}
\newcommand{\mrR}{{\mathrm{R}}}
\newcommand{\mrS}{{\mathrm{S}}}
\newcommand{\mrT}{{\mathrm{T}}}
\newcommand{\mrU}{{\mathrm{U}}}
\newcommand{\mrZ}{{\mathrm{Z}}}
\newcommand{\ssOS}{{\mathrm{OS}}}
\newcommand{\mcA}{\mathcal{A}}
\newcommand{\mcD}{\mathcal{D}}
\newcommand{\mcL}{\mathcal{L}}
\newcommand{\mcO}{\mathcal{O}}
\newcommand{\mcP}{\mathcal{P}}
\newcommand{\mcZ}{\mathcal{Z}}
\newcommand{\mrdim}{\mathrm{dim}}
\newcommand{\DUV}{\Delta_{\mathrm{UV}}}
\newcommand{\emc}{\mathswitch \gamma}
\newcommand{\muR}{\mathswitch {\mu_{\mrR}}}
\newcommand{\muRs}{\mathswitch {\mu^2_{\mrR}}}
              
\newcommand{\ctw}{\mrc_{_{\theta}}}
\newcommand{\stws}{\mrs_{_{\theta}}^2}
\newcommand{\ctws}{\mrc_{_{\theta}}^2}
\newcommand{\mw}{\mathswitch{M_{\PW}}}
\newcommand{\mz}{\mathswitch{M_{\PZ}}}
\newcommand{\mws}{\mathswitch{M^2_{\PW}}}

\newcommand{\eg}{e.g.\xspace}

\newcommand{\etc}{etc.\@\xspace}
\DeclareRobustCommand{\PA}{\HepParticle{A}{}{}\Xspace}
\DeclareRobustCommand{\PV}{\HepParticle{V}{}{}\Xspace}

\newcommand{\sPW}{{\scriptscriptstyle{\PW}}}
\DeclareRobustCommand{\PXpm}{\HepParticle{\PX}{}{\pm}\Xspace}
\DeclareRobustCommand{\PX}{\HepParticle{X}{}{}\Xspace}

\newcommand{\sPHZZ}{{\scriptscriptstyle{\PH\PZ\PZ}}}
\newcommand{\sPHWW}{{\scriptscriptstyle{\PH\PW\PW}}}
\newcommand{\Pgg}{\Pg\Pg}
\newcommand{\sPgg}{{\scriptscriptstyle{\Pgg}}}
\newcommand{\ren}{{\mbox{\scriptsize ren}}}
\newcommand{\fin}{{\mbox{\scriptsize fin}}}
\newcommand{\myexp}{{\mbox{\scriptsize exp}}}
\newcommand{\EFT}{\rm{\scriptscriptstyle{SMEFT}}}

\newcommand{\myLO}{\rm{\scriptscriptstyle{LO}}}
\newcommand{\myNLO}{\rm{\scriptscriptstyle{NLO}}}
\newcommand{\myQED}{\rm{\scriptscriptstyle{QED}}}
\newcommand{\mySM}{\rm{\scriptscriptstyle{SM}}}

\newcommand{\srt}{\sqrt{2}}
\newcommand{\PWW}{\PW\PW}

\newcommand{\sPA}{{\scriptscriptstyle{\PA}}}
\newcommand{\ctwr}{\mrc^{\ren}_{_{\theta}}}

\newcommand{\mzOS}{\mathswitch{M_{\PZ\,;\,\ssOS}}}
\newcommand{\mzsOS}{\mathswitch{M^2_{\PZ\,;\,\ssOS}}}
\DeclareRobustCommand{\PK}{\HepParticle{\upPhi}{}{}\Xspace}

\DeclareRobustCommand{\Pf}{\HepParticle{f}{}{}\Xspace}

\newcommand{\PVV}{\PV\PV}
\newcommand{\PAA}{\PA\PA}
\newcommand{\cth}{{\hat \mrc}_{_{\theta}}}

\newcommand{\cths}{{\hat \mrc}^2_{_{\theta}}}

\newcommand{\sths}{{\hat \mrs}^2_{_{\theta}}}

\newcommand{\nfct}{{\mbox{\scriptsize nfc}}}
\def\nn{\nonumber\\ }

\def\lsix{ \mathcal{L}^{(6)}}

\def\hyp{\mathsf{y}}

\def\lsix{ \mathcal{L}^{(6)}}
\def\gcb{{\overline g_{1}}}
\def\gcw{{\overline g_{2}}}
\def\gcg{{\overline g_{3}}}
\def\gcZ{{\overline g_{Z}}}

\def\tc{{\overline \theta}}
\def\ec{{\overline e}}
\def\sc{{\overline s}}

\def\ckin{c_{H,\text{kin}}}
\def\that{{\hat \theta}}